\documentclass[journal]{IEEEtran}
\IEEEoverridecommandlockouts
\usepackage{url}
\usepackage{cite}
\usepackage{amsmath,amssymb,amsfonts}
\usepackage{algpseudocode}
\usepackage{algorithm}
\usepackage{multirow}
\usepackage{threeparttable}
\usepackage{booktabs}
\makeatletter
\newcommand\fs@betterruled{%
  \def\@fs@cfont{\bfseries}\let\@fs@capt\floatc@ruled
  \def\@fs@pre{\vspace*{8pt}\hrule height.8pt depth0pt \kern2pt}%
  \def\@fs@post{\kern2pt\hrule\relax}%
  \def\@fs@mid{\kern2pt\hrule\kern2pt}%
  \let\@fs@iftopcapt\iftrue}
\floatstyle{betterruled}
\restylefloat{algorithm}
\makeatother
\usepackage{graphicx}
\usepackage{textcomp}
\usepackage{xcolor}
\def\BibTeX{{\rm B\kern-.05em{\sc i\kern-.025em b}\kern-.08em
    T\kern-.1667em\lower.7ex\hbox{E}\kern-.125emX}}
\newtheorem{remark}{Remark}
\usepackage{mathrsfs}
\ifCLASSOPTIONcompsoc
    \usepackage[caption=false, font=normalsize, labelfont=sf, textfont=sf]{subfig}
\else
\usepackage[caption=false, font=footnotesize]{subfig}
\fi
\usepackage{mathtools}
\begin{document}

\title{Resilient and Reliable Cloud Network Control for Mission-Critical Latency-Sensitive Service Chains}
\author{Chin-Wei Huang,~\IEEEmembership{Graduate~Student~Member,~IEEE,}
        Jaime Llorca,~\IEEEmembership{Senior~Member,~IEEE,}\\
        Antonia M. Tulino,~\IEEEmembership{Fellow,~IEEE,}
        and Andreas F. Molisch,~\IEEEmembership{Fellow,~IEEE}}
\maketitle

\begin{abstract}
The proliferation of mission-critical latency-sensitive services has intensified the demand for next-generation cloud-integrated networks to guarantee both {\em reliable} and {\em resilient} service delivery. While reliability imposes timely-throughput requirements, i.e., percentage of packets to be delivered within a prescribed per-packet deadline, resilience relates to the network's ability to swiftly recover timely-throughput performance following an outage event, such as node or link failures.
While recent studies have increasingly focused on designing reliable network control policies, a comprehensive solution that combines reliable and resilient network control has yet to be fully explored. 
This paper formulates the {\em multi-commodity least-cost resilient and reliable network control (MC-LC-ResRNC)} problem as a stochastic control problem with long and short-term timely throughput constraints. 
We then present a solution through the Multi-Commodity Resilient and Reliable Cloud Network Control (MC-ResRCNC) algorithm and show through numerical experiments that it jointly ensures reliability under normal conditions and resilience upon network failure. 
\end{abstract}

\begin{IEEEkeywords}
Cloud network control, service function chaining, end-to-end (E2E) delay constraint, packet deadline, timely throughput, reliability, resilience, outage, network failure, Lyapunov optimization
\end{IEEEkeywords}

\section{Introduction}
\IEEEPARstart{T}{th} rapid evolution of next-generation communication infrastructure is paving the way for cloud-integrated and edge-empowered networks that can support emerging classes of mission-critical and latency-sensitive services.
These include industrial automation, autonomous vehicles (AVs), remote surgery, extended reality (XR), Internet of Drones (IoD), and critical infrastructure monitoring---applications where per-packet timeliness and continuity of service directly determine system functionality and safety.

As computation becomes increasingly distributed across the network (via e.g., distributed cloud, edge, and fog resources), such services rely on tightly coordinated joint communication-computation control to meet stringent end-to-end (E2E) latency guarantees and sustained network throughput.
%As these networked services become deeply intertwined with the physical world, a wide range of mission-critical latency-sensitive applications require guaranteeing consistent fulfillment of stringent latency requirements across communication and computation domains becomes paramount. 
In industrial control and cyber-physical systems, sensor data must be processed within a few milliseconds to maintain safety \cite{5gacia2019automation}; autonomous vehicles require sub-100 ms E2E latency for cooperative perception \cite{3gpp2020ts22185, 5gamericas2019}; and interactive XR or telesurgery demands $\leq$20 ms motion-to-photon (MTP) latency\cite{etsi2021tr126928, ericsson2021xr} and $\leq$100 ms delay for telesurgery\cite{nankaku2022telesurgerydelay} to ensure perceptual continuity and safe operation.
A defining feature of these mission-critical applications is the need to sustain the throughput while satisfying strict {\em per-packet} E2E deadlines---the timely throughput.
Specifically, the timely throughput is the rate of packets delivered to their destinations before they expire \cite{RCNC, Vitale2025Delay, Fountoulakis2023Scheduling, Singh2019Throughput, Chen2018Timely}\footnote{Instead of using timely throughput directly as an optimization objective or imposing a direct constraint on it, it is the \emph{ratio} of timely throughput to the total arrival rate that is generally used as the metric to evaluate the reliable transmission for mission-critical applications.}.
In this context, two key requirements are involved in addition to optimizing cost-efficiency in a cloud network control policy: 
i) \textbf{Reliability}---the assurance of timely throughput over both long and short intervals under stationary conditions;
ii) \textbf{Resilience}---the capability to recover and maintain timely throughput after non-stationary events such as node or link failures.
Together, reliability and resilience form the foundation of robust network control, a paradigm essential for next-generation distributed cloud networks supporting mission-critical and latency-sensitive applications.

As the importance of reliability and resilience has been widely recognized, they have been investigated in cloud-network control and performance analysis, though focusing primarily on {\em either} reliability {\em or} resilience. 
Studies emphasizing reliability without resilience include \cite{RCNC, Pagliuca2024Dual, Singh2022Energy, Lashgari2021End, Vitale2025Delay, Xia2014Distributed}.
The impacts of network outages and corresponding failure detection or recovery mechanisms have also been extensively studied \cite{Andersen2001Resilient, Cetinkaya2013Modeling, STERBENZ20101245, Li2025Achieving, Verma2025Toward}.
Some works consider both reliability and resilience, yet rely on simplified delay models rather than strict per-packet E2E latency constraints \cite{Barla2013Optimal, Qu2017A, Mogyorosi2022Resilient, Babarczi2023Intelligent, Zhang2019Tripod, Li2023Estimating, Wenhao20245G, Nouruzi2025AI, Mahmood2025Resilient}. 
In particular, Ref. \cite{Barla2013Optimal} proposes resilience models that minimize cost and propagation delay but without per-packet delay constraint, while Refs. \cite{Qu2017A, Mogyorosi2022Resilient, Babarczi2023Intelligent} impose hop-count limitations rather than actual delay guarantees for their algorithm with awareness of reliability and resilience.
Ref. \cite{Zhang2019Tripod} analyzes per-packet E2E delay for their high-performance and failure-resilience framework, but does not enforce it as a constraint, and Refs. \cite{Li2023Estimating, Wenhao20245G} estimate resilience performance without developing a control algorithm.
Ref. \cite{Nouruzi2025AI} introduces a deep reinforcement learning-based approach for resilient cost minimization under network slicing, yet constrains latency by the total delay per user rather than per packet.
Ref. \cite{Mahmood2025Resilient}  proposed a ``Resilient-by-Design'' framework, outlining a four-stage process---predict, preempt, protect, and progress---to ensure that 6G systems can withstand disruptions and recover rapidly, but no latency constraint is included.

In summary, although prior works underscore the importance of reliability and resilience, {\em no network-control algorithm has jointly addressed both under strict per-packet E2E latency constraints}---a requirement vital for ensuring the high performance and availability demanded by mission-critical applications.
To bridge this gap, this paper proposes {\em a cloud network control algorithm that jointly minimizes transmission and processing costs while ensuring resilience and reliability under strict per-packet latency constraints}.

This paper thus makes the following key contributions:
\begin{enumerate}
    \item \textbf{Joint Reliable and Resilient Network Control}: We define a comprehensive network control problem---the Multi-Commodity Least-Cost Resilient and Reliable Network Control (MC-LC-ResRNC) problem---addressing the need for both reliable service delivery under stationary conditions and resilient recovery in response to outages and non-stationary network changes. 
    The problem is formulated to minimize both transmission and processing costs while meeting E2E latency, reliability, and resilience constraints.
    \item \textbf{Development of the MC-ResRCNC Algorithm}: 
    We propose the Multi-Commodity Resilient and Reliable Cloud Network Control (MC-ResRCNC) algorithm, which is designed to adaptively manage network computation and communication resources across multiple service chains to ensure reliable throughput in steady-state conditions and resilient recovery after non-stationary events. 
    The MC-ResRCNC algorithm leverages Lyapunov drift-plus-penalty methods and flow matching techniques over a layered-graph topology to optimize routing and scheduling decisions for multiple commodities in a service chain.
    \item \textbf{Performance Analysis through Numerical Experiments}: Through extensive simulations, we validate the effectiveness of the MC-ResRCNC algorithm in providing reliability and resilience. 
    Our results demonstrate that MC-ResRCNC significantly outperforms the state-of-the-art by achieving much faster timely throughput recovery following outages and more consistent performance under steady-state conditions.
    Further analysis of the behavior of MC-ResRCNC after an outage demonstrates the necessity and a possible strategy for reducing service arrival rates after an outage.
\end{enumerate}

\section{System Model}
We consider a setting in which a set of multi-commodity service chains is hosted by a distributed computing network (e.g., cloud-integrated network, edge computing network).

%We model the network as a directed graph with nodes representing network devices with packet queuing capabilities and, in some cases, also data processing capabilities, and edges representing communication links between network nodes. 
%Network services are described as multi-commodity chains, where the packet transmission and processing are jointly described with the layered graph, an extension from the directed graph \cite{layered_graph}.
%To assess the resilience of these systems, we account for failures or outage events by removing the affected nodes and links.
%These outages introduce non-stationary conditions, challenging the network to adapt and recover timely throughput under constrained configurations.

\subsection{Service Model}
\begin{figure}[t]
     \centering 
        \includegraphics[width=1\linewidth]{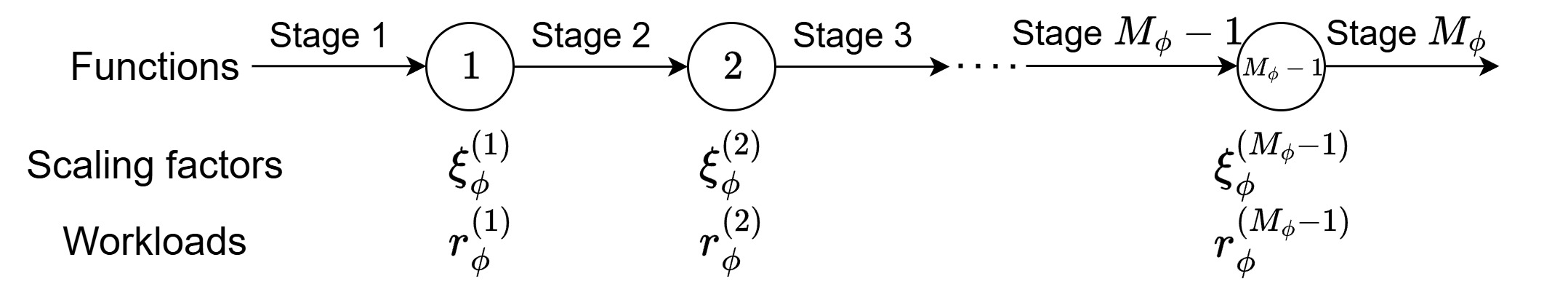}
     \caption{Service function chain of service $\phi$}
\label{fig:service_function_chain}
     \vspace{-15pt}
\end{figure}
We describe a network service $\phi\in\Phi$ by a service function chain \cite{layered_graph} with $M_{\phi}$ stages. 
We denote by $\mathcal{M}_{\phi} =\{1, 2, \dots, M_{\phi}\}$ the ordered set of its stages, which implies the existence of $M_{\phi}-1$ functions in between.
Each function is described by its scaling factor and workload, %in the unit of [packet/packet] and [CPU $\cdot$ time slot/packet], 
which represent the number of generated output packets per input packet and the required processing resources (e.g., number of operations or FLOPs) needed per input packet, respectively.
We denote by $\xi_{\phi}^{(m)}$ and $r_{\phi}^{(m)}$ the scaling factor and workload of the $m$-th function in service $\phi$, respectively.
The service function chain structure is illustrated in Fig. \ref{fig:service_function_chain}, in which the numbered circles, which represent the service functions, are connected in order by the arrows, which represent the corresponding processing stages. %, together to form the service function chain of service $\phi\in\Phi$.

% While for ease of presentation, we focus on a single service chain, all models and algorithms in this paper can be readily extended to multiple service chains.

\subsection{Network Model}\label{subsec:network_model}
The network is modeled by the directed graph $\mathcal{G}=(\mathcal{V}, \mathcal{E})$, where node $i\in\mathcal{V}$ represents a network device (e.g., cloud server, edge server, compute-enabled access point, end-user device) and edge $(i,j)\in\mathcal{E}$ represents a link supporting data transmission from node $i$ to node $j$.

As in \cite{layered_graph}, packet transmission and processing are jointly described via a {\bf layered graph transformation} of $\mathcal G$, where intra-layer links represent communication resources and inter-layer links computation resources.
%\subsection{Layered Graph Extension for Packet Processing}\label{subsec:layered_graph}
%To jointly include the processing, i.e., the functions, in the services with the packet transmission in the network control policy, we adopt the layered graph structure \cite{layered_graph} to extend the directed graph for each service. 
In particular, 
for service $\phi$, its layered graph $\mathcal{G}^{(\phi)}$ has $M_{\phi}$ layers, in which each layer is a copy of the original directed graph $\mathcal{G}$, and the node $i\in\mathcal{V}$ on the $m$-th layer is denoted by $i_{m}$. 
While the $M_{\phi}$ layers accommodate the packets at the $M_{\phi}$ stages in service $\phi$, the adjacent layers are connected by a directed link on every node, which represents the processing link between the corresponding two stages (i.e., the function between them) at that node. 
Specifically, the transmission of the packets at the $m$-th stage from node $i$ to node $j$ is represented by link $(i_{m},j_{m})$, and the $m$-th function done at node $i$ is represented by link $(i_{m},i_{m+1})$.
In summary, intra-layer transmission refers to the packet transmission at the corresponding stage within the network, and the inter-layer transmission represents the packet processing of the corresponding function at the corresponding node.
To sum up, we define $\mathcal{G}^{(\phi)}=(\mathcal{V}^{(\phi)}, \mathcal{E}^{(\phi)})$ with 
\begin{align}
    \mathcal{V}^{(\phi)} &= \{i_m:i\in\mathcal{V}, 1\leq m\leq M_{\phi}\}\\
    \mathcal{E}^{(\phi)} &= \left(\cup_{i\in \mathcal{V}}\,\mathcal{E}^{(\phi)}_{\mathrm{pr}, i}\right)\cup \left(\cup_{(i,j)\in \mathcal{E}}\,\mathcal{E}^{(\phi)}_{\mathrm{tr}, ij}\right)
\end{align}
where $\mathcal{E}^{(\phi)}_{\mathrm{pr}, i}=\{(i_m, i_{m+1}): 1\leq m\leq M_{\phi}-1\}$ is the set of processing links on node $i$ for all functions in service $\phi$, and $\mathcal{E}^{(\phi)}_{\mathrm{tr}, ij}=\{(i_m, j_{m}): 1\leq m\leq M_{\phi}\}$ is the set of transmission links on link $(i,j)$ at all stages in service $\phi$.
For any node $\imath\in\mathcal{V}^{(\phi)}$, we further define $\delta_{\imath}^{(\phi,+)}=\{\jmath\in \mathcal{V}^{(\phi)}: (\imath,\jmath)\in \mathcal{E}^{(\phi)}\}$ and $\delta_{\imath}^{(\phi,-)}=\{\jmath\in \mathcal{V}^{(\phi)}: (\jmath,\imath)\in \mathcal{E}^{(\phi)}\}$, which are the set of its outgoing and incoming neighbor nodes, respectively.
Also, we define the following auxiliary variables for the purpose of presentation:
\begin{align}
(\zeta_{\imath\jmath}^{(\phi)}, \rho_{\imath\jmath}^{(\phi)})=
\begin{cases}
    (1,1) & \text{if } (\imath,\jmath)=(i_m,j_{m})\\
    (\xi_{\phi}^{(m)}, r_{\phi}^{(m)}) & \text{if }(\imath,\jmath)=(i_m,i_{m+1}).
\end{cases}
\end{align}
for any $i\in\mathcal{V}$, $(i,j)\in\mathcal{E}$, $m\in\mathcal{M}_{\phi}$, and $\phi\in\Phi$.\footnote{The $\zeta_{\imath\jmath}^{(\phi)}$ and $\rho_{\imath\jmath}^{(\phi)}$ in the transmission case are unitless numbers that help the writing of formulas without an actual meaning for its multiplication or division with the transmission flow rate or capacity.} 
Note that, because the exogenous packets arrive to undergo all the functions, a source node $s\in\mathcal{V}$ and a destination node $d\in\mathcal{V}$ correspond to $s_{1}$ and $d_{M_{\phi}}$ in the layered graph, which are on the first and the last layer, respectively.

Time is divided into equal-sized slots, whose duration can be chosen in practice according to network structure and service requirements.
A packet transmission (resp. packet processing) over a communication link (resp. computation link) is assumed to take one time slot.
The capacity $C_{ij}$, in the unit of packet per time slot, denotes the maximum number of data packets that can be delivered (transmitted or processed) in one time slot over link $(i,j)$, and the cost $e_{ij}$, in the unit of cost per packet, the cost of delivering (transmitting or processing) a data packet in one time slot over link $(i,j)$.

\subsection{Traffic Model}
% \textcolor{blue}{
% \begin{itemize}
%     \item destination node
%     \item arrival process with lifetime
%     \item timely throughput
% \end{itemize}
% }
A commodity $c\in \mathcal{C}$ is specified by its destination node $d_{c}$ and corresponding service $\phi_{c}\in\Phi$.
A commodity $c$ packet, which can have an exogenous arrival at any source node in the network, should undergo the $M_{\phi_c}-1$ functions in order before being consumed at the destination node $d^{(c)}$.
Such a packet has a given \emph{lifetime} or \emph{time-to-live} (TTL) $l\in\mathcal{L}_{a}^{(c)}=\{L_{\mathrm{min}}^{(c)}, L_{\mathrm{min}}^{(c)} + 1, \dots, L_{\mathrm{max}}^{(c)}\}$, defined as the number of time slots the packet can stay in the network before becoming useless in the latency-sensitive application.
The arrival of each exogenous packet is modeled by a random process, namely the arrival process. 
Specifically, we denote the number of exogenous commodity $c$ packets that arrive in time slot $t$ with lifetime $l\in\mathcal{L}_a^{(c)}$ at node $i$ by $a^{(c,l)}_{i}(t)$, which is a nonnegative-valued stationary random process with time-independent probability density function (pdf) $\boldsymbol{f}_{i}^{(c, l)}$ in each time slot. 
For clarity, we further define vector $\boldsymbol{a}^{(c)}(t)\in\mathbb{R}_+^{|\mathcal{L}_a^{(c)}|\cdot|\mathcal{V}|.}$, which includes $a^{(c,l)}_{i}(t)$ for all $l\in \mathcal{L}_a^{(c)}$ and all $i\in\mathcal{V}$.
With the lifetime in packets, the timely throughput is further defined as the average number of packets per time slot that are successfully delivered (before expiration) to the destination.

\subsection{Queuing System}
% \textcolor{blue}{
% \begin{itemize}
%     \item commodity and lifetime-based queue backlog 
%     \item commodity and lifetime-based flow rate 
%     \item queuing dynamic
%     \item queue expiration and consumption
% \end{itemize}
% }
To keep track of packets with their corresponding lifetime and commodity, we denote by $Q_{\imath}^{(c,l)}(t)$ the number of queued commodity $c$ packets with lifetime $l$ at node $\imath\in\mathcal{V}^{(\phi_c)}$, which are at the $m$-th stages in service $\phi_c$ at node $i\in\mathcal{V}$ if $\imath = i_m$.
Also, we denote by $x_{\imath\jmath}^{(c,l)}(t)$ the number of commodity $c$ packets with lifetime $l$ transmitted from node $\imath$ to node $\jmath$ per time slot at time $t$.\footnote{The ``transmission'' here can mean either packet transmission or processing according to whether the link $(\imath,\jmath)$ is in a set of transmission links or a set of processing links.}
The queuing dynamics are given by
\begin{equation}
\begin{aligned}
    & Q_{\imath}^{(c,l)}(t+1) = Q_{\imath}^{(c,l+1)}(t) - x_{\imath \to}^{(c,l+1)}(t) + x_{\to \imath}^{(c,l+1)}(t)\\
    &\quad  + a_{\imath}^{(c,l)}(t+1), \; \forall c\in \mathcal{C}, l\in \mathcal{L}^{(c)}
\end{aligned}\label{queuing_dynamics}
\end{equation}
where $x_{\imath\to}^{(c, l+1)}(t)=\sum_{\jmath\in\delta_{\imath}^{(\phi_c,+)}}x_{\imath\jmath}^{(c, l+1)}(t)$ and $x_{\to\imath}^{(c, l+1)}(t)=\sum_{\jmath\in\delta_{\imath}^{(\phi_c,-)}}x_{\jmath\imath}^{(c, l+1)}(t)$ are the number of outgoing and incoming packets per time slot at node $\imath$, respectively\footnote{Flow rate $x_{\imath\jmath}^{(c,L_{\mathrm{max}}^{(c)}+1)}(t)$ and queue backlog $Q_{\imath}^{(c,L_{\mathrm{max}}^{(c)}+1)}(t)$ are always $0$ for all $(\imath, \jmath)\in \mathcal{E}^{(\phi_c)}$ and $\imath\in \mathcal{V}^{(\phi_c)}$ because the assumption of maximal lifetime $L_{\mathrm{max}}^{(c)}$.}, $\mathcal{L}^{(c)}=\{1,2, \dots, L^{(c)}_{\mathrm{max}}\}$ is the set of all possible lifetimes of a commodity $c$ packet, and 
\begin{equation}
    a_{\imath}^{(c,l)}(t+1) = \begin{cases}
        a_{i}^{(c,l)}(t+1),  & \text{if }\imath =i_1\\
        0 & \text{otherwise}.
    \end{cases}
\end{equation}
Furthermore, because outdated packets should be dropped and packets arriving at destination $d^{(c)}_{M_{\phi_c}}\in\mathcal{V}^{(\phi_c)}$ are consumed immediately, the queues also have the following properties:
\begin{equation}
\begin{aligned}
    &Q_{\imath}^{(c, 0)}(t)=Q_{d^{(c)}_{M_{\phi_c}}}^{(c, l)}(t)=0, \; \forall \imath\in\mathcal{V}^{(\phi_c)},  l\in\mathcal{L}^{(c)}.
\end{aligned}\label{pkt_expiration_and_consumption}
\end{equation}

\subsection{Outage Event}\label{sec:outage_event}
% \textcolor{blue}{
% \begin{itemize}
%     \item outage time $t_{o}$
%     \item directed graph after outage $\mathcal{G}^{(\phi)}_{o}$
%     \item arrival rate before/after outage 
% \end{itemize}
% }
The network system in regular operation can be described with the system model presented above. 
However, the change of topology and corresponding configurations due to an outage event needs to be considered separately. 
Hence, we model the physical network topology after an outage by the directed graph $\mathcal{G}_{o}=\{\mathcal{V}_o, \mathcal{E}_o\}$, which is obtained by removing the nonfunctional links and nodes in $\mathcal{G}$.
The corresponding layered graph for each service are denoted by $\mathcal{G}_{o}^{(\phi)}=\{\mathcal{V}_o^{(\phi)}, \mathcal{E}_o^{(\phi)}\}$, which is build directly upon $\mathcal{G}_{o}$ accordingly to Sec. \ref{subsec:network_model}. 
The arrival process is also assumed to be different in terms of its mean after the outage time $t_o$ to adapt to the limitations imposed by the new topology after the outage event. 
Therefore, we have mean arrival rate before outage denoted by $\lambda_{i}^{(c,l)}\triangleq \mathbb{E}\{a_i^{(c,l)}(t)\}$, for all $t < t_{o}$ and mean arrival rate after outage denoted by  $\lambda_{i, o}^{(c,l)}\triangleq \mathbb{E}\{a_i^{(c,l)}(t)\}$, for all $ t\geq t_o$.
Typically, $\lambda_{i, o}^{(c,l)}$ is less than or equal to $\lambda_{i}^{(c,l)}$ in response to the possibly reduced ability to support reliable transmission. 
We also define vectors $\boldsymbol{\lambda}^{(c)}=\mathbb{E}\{\boldsymbol{a}^{(c)}(t)\}$, for all $t < t_{o}$ and $\boldsymbol{\lambda}_{o}^{(c)}=\mathbb{E}\{\boldsymbol{a}^{(c)}(t)\}$, for all $t \geq t_{o}$.
% Vectors 
% $\boldsymbol{\lambda}=[(\boldsymbol{\lambda}^{(1)})^T\dots (\boldsymbol{\lambda}^{(|\mathcal{C}|)})^T]^T$ and $\boldsymbol{\lambda}_o=[(\boldsymbol{\lambda}^{(1)}_o)^T\dots (\boldsymbol{\lambda}^{(|\mathcal{C}|)}_o)^T]^T$ are also defined for presentation.

\section{Problem Formulation}\label{sec:prob_formulation}
Reliable and resilient delivery of mission-critical, latency-sensitive applications requires the timely throughput to be above a certain quality-of-service (QoS) level on both long-term average (as $t$ grows to infinity) and short-term average (over short time periods).
In the design of a resilient system, an outage event can lead to a temporary increase in expired packets, resulting in a lower timely throughput for a short time period without affecting the long-term timely throughput. 
Hence, the ability and speed of recovering a satisfactory timely throughput after an outage event, a key resilience property, cannot be evaluated through a long-term average metric.
In the following, we provide formal definitions for long and short-term timely throughput metrics, associated reliability and resilience constraints, and overall problem formulation.

\subsection{Reliability and Resilience Constraints}\label{subsec:reliability}
We first define the {\em short-term timely throughput} of commodity $c$ as its time-average of the packets delivered to the destination over time window $[t,t+T-1]$ of duration $T$:
\vspace{-2mm}
\begin{equation}
    R^{(c)}(t, T) = \frac{1}{T}\sum_{\tau = t}^{t+T-1} x_{\to d_{c}}^{(c)}(\tau),
\label{stthroughput}
\end{equation}
where $x^{(c)}_{\to d_{c}}(\tau) =\sum_{l\in\mathcal{L}^{(c)}}\sum_{\imath\in\delta_{d_{c}}^{(\phi_c,-)}}x_{\imath d_{c}}^{(c, l)}(\tau)$ is the timely throughput at time $\tau$ and $d_{c}=d_{M_{\phi_c}}^{(c)}$ is the destination node on the corresponding layered graph. 
Similarly, its {\em short-term arrival rate} is defined as 
\vspace{-1mm}
\begin{equation}
    A^{(c)}(t, T) = \frac{1}{T}\sum_{\tau = t}^{t+T-1} \|\boldsymbol{a}^{(c)}(\tau)\|_1.
    \label{starrivalrate}
\end{equation}
From \eqref{stthroughput} and \eqref{starrivalrate},  the {\em long-term timely throughput} and {\em long-term arrival rate} of commodity $c$ can be defined as $\lim_{T\rightarrow \infty }R^{(c)}(0, T)$ and $\lim_{T\rightarrow \infty }A^{(c)}(0, T)$, respectively.
Then, the \emph{short-term reliability level} is defined as
\begin{equation}\label{short-term_reliability_level}
     \frac{R^{(c)}(t, T)}{ \Xi_{\phi_{c}}^{(M_{\phi_c})}A^{(c)}(t, T)}, 
\end{equation}
where $\Xi_{\phi}^{(m)}\triangleq \prod_{s=1}^{m-1}\xi_{\phi}^{(s)}$ for $2\leq m\leq M_{\phi}$ and $\Xi_{\phi}^{(1)}\triangleq 1$ for any service $\phi$ are the overall scaling factors.
The \emph{long-term reliability level} is further defined by \eqref{short-term_reliability_level} with $t=0$ and $T\to \infty$ separately for the numerator and denominator.

We can now introduce the {\em long-term} {\bf reliability constraint}, which imposes the long-term timely throughput to be above a prescribed fraction $\gamma_{\mathrm{long}}^{(c)}\in [0,1]$ (we call it \emph{long-term reliability level requirement}) of the long-term arrival rate in their expectation:
\begin{equation}
    \lim_{T\rightarrow \infty }\mathbb{E}\{R^{(c)}(0, T)\} \geq \gamma_{\mathrm{long}}^{(c)} \Xi_{\phi_{c}}^{(M_{\phi_c})}\lim_{T\rightarrow \infty }\mathbb{E}\{A^{(c)}(0, T)\},
    \label{reliability_constraint}
\end{equation}
which also implies that the long-term reliability level is larger than or equal to its required value $\gamma_{\mathrm{long}}^{(c)}$ in terms of the ratio of expectations.
\begin{figure}[t]
    \vspace{3pt}
    \centering \includegraphics[width=0.68\linewidth]{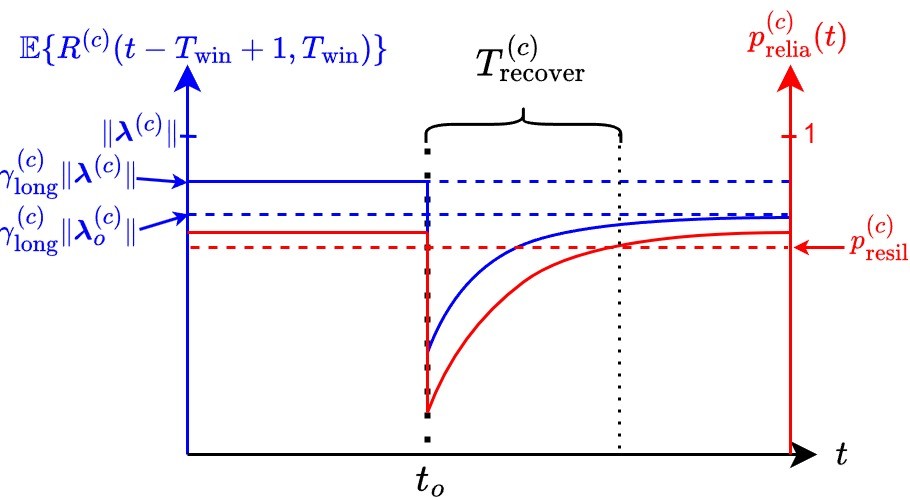}
    \vspace{-8pt}
    \caption{Illustration of the short-term timely throughput and the reliability probability evolution before and after an outage event.}
    \label{fig:throughput_change}
    \vspace{-16pt}
\end{figure}

In addition to reliability, resilient service delivery requires the network to recover a prescribed level of timely throughput after a finite time window following an outage event. 
To formalize this, we define the \emph{short-term reliability satisfaction event} as the event that the short-term timely throughput, measured over the interval \([t - T_{\mathrm{win}} + 1, t]\), exceeds a prescribed fraction \(\gamma_{\mathrm{short}}^{(c)}\in[0,1]\) (we call it \emph{short-term reliability level requirement}) of the short-term arrival rate, also measured over the same interval:
\begin{equation}
R^{(c)}(t - T_{\mathrm{win}}+1, T_{\mathrm{win}})\geq \gamma_{\mathrm{short}}^{(c)}\Xi_{\phi_{c}}^{(M_{\phi_c})}A^{(c)}(t - T_{\mathrm{win}}+1, T_{\mathrm{win}})\label{short-term_reliability}
\end{equation}
which also implies that the corresponding short-term reliability level is equal to or larger than its required value $\gamma_{\mathrm{short}}^{(c)}$.
Then, we denote by $p_{\mathrm{relia}}^{(c)}(t)$ the probability of this event, which we call the \emph{short-term reliability satisfaction probability}. 
The {\bf resilience constraint} requires that the short-term reliability satisfaction probability to be not below the resilience probability requirement $p_{\mathrm{resil}}^{(c)}$ for all time slots after a prescribed recovery time \(T_{\mathrm{recover}}^{(c)}\) following the outage time \(t_o\), that is, 
\begin{equation}
      p^{(c)}_{\mathrm{relia}}(t)\geq p^{(c)}_{\mathrm{resil}}, \,\forall t\in\mathcal{T}_{\mathrm{resil}}^{(c)}=\{t\in\mathbb{N}:t\geq t_{\mathrm{recover}}^{(c)}\}, 
    \label{resilience_constraint}
\end{equation}
where $t^{(c)}_{\mathrm{recover}}=t_o+T^{(c)}_{\mathrm{recover}}$.
A policy with such recovery ability described in \eqref{resilience_constraint} can be seen in an illustrative example in Fig. \ref{fig:throughput_change}, where the probability of the short-term timely throughput $p_{\mathrm{relia}}^{(c)}(t)$ as well as its expectation $\mathbb{E}\{R^{(c)}(t-T_{\mathrm{win}}+1, T_{\mathrm{win}})\}$ gradually recover to the required levels $p_{\mathrm{resil}}^{(c)}$ and $\gamma_{\mathrm{long}}^{(c)}\|\boldsymbol{\lambda}_o^{(c)}\|$ from \eqref{reliability_constraint} and \eqref{resilience_constraint}, respectively, after the outage time $t_{o}$.

\subsection{Admissible Policy Space}
The control policies of interest make packet routing, scheduling, and processing decisions at every time slot, which is described jointly by the flow rate $x_{\imath\jmath}^{(c, l)}(t)$ across the layered graphs of all the services. 
Therefore, the space of admissible control policies can be defined as the region satisfying:
\begin{enumerate}
    \item non-negativity constraint, i.e.,
    \begin{equation}
        x^{(c, l)}_{\imath \jmath}(t) \geq 0, \;\; \forall (\imath,\jmath)\in\mathcal{E}^{(\phi_c)}, l\in\mathcal{L}^{(c)}, c\in\mathcal{C},\label{nonnegativity_constraint}
    \end{equation}
    \item transmission and processing capacity constraint, i.e.,
    \begin{subequations}\label{capacity_constraints}
        \begin{flalign}
           &\sum_{c\in\mathcal{C}} \sum_{(\imath, \jmath)\in \mathcal{E}^{(\phi_c)}_{\mathrm{tr},ij}}\sum_{l\in\mathcal{L}^{(c)}}x^{(c, l)}_{\imath \jmath}(t) \leq C_{ij}, \forall (i,j)\in\mathcal{E},\label{transmission_capacity_constraint}\\
            &\sum_{c\in\mathcal{C}}\sum_{(\imath, \jmath)\in \mathcal{E}^{(\phi_c)}_{\mathrm{pr},i}}\sum_{l\in\mathcal{L}^{(c)}}\rho_{\imath\jmath}^{(\phi_c)}x^{(c, l)}_{\imath \jmath}(t) \leq C_{i}, \forall i\in\mathcal{V},\label{processing_capacity_constraint}
        \end{flalign}
    \end{subequations}
    \item availability constraint, i.e., 
    \begin{equation}
        x^{(c, l)}_{\imath\rightarrow}(t) \leq Q^{(c, l)}_\imath(t), \;\;\forall \imath\in\mathcal{V}^{(\phi_c)}, l\in\mathcal{L}^{(c)}, c\in\mathcal{C},\label{availability_constraint}
    \end{equation}
    where $x_{\imath \to}^{(c,l)}(t)=\sum_{\jmath\in\delta_{\imath}^{(\phi_c,+)}} x_{\imath \jmath}^{(c,l)}(t)$.
\end{enumerate}

\subsection{Reliability Region} 
In \cite{RCNC}, the stability region is defined as the region where there exists an admissible policy that satisfies the reliability constraint. In this paper, to evaluate the reliability of a specific control policy, we define the reliability region as the set of parameters $(\{\boldsymbol{f}_{i}^{(c, l)}, \boldsymbol{f}_{i,o}^{(c, l)}\}_{i\in\mathcal{V}, c\in\mathcal{C},l\in \mathcal{L}}, \gamma_{\mathrm{long}})$ such that the corresponding admissible policy can satisfy the reliability constraint (i.e., satisfying \eqref{nonnegativity_constraint}, \eqref{capacity_constraints}, \eqref{availability_constraint}, and \eqref{reliability_constraint} for all $c\in\mathcal{C}$). 
To evaluate the reliability region in practice, the $\lim_{T\to \infty}\mathbb{E}\{R(0, T)\}$ in \eqref{reliability_constraint} is separately estimated by taking the sample mean of $R(0,t_{o})$ and $R(t_{o}, T)$ with $T$ being the number of time slots after outage in the simulation for the reliability region before and after the outage event, respectively.\footnote{The number of time slots before and after the outage time $t_{o}$ should be large enough in the simulation to ensure a negligible estimation error.}

\subsection{Resilience Region}\label{sec:res_region}
To analyze the resilience of a specific network control policy on a given network with certain requested services, the resilience region is defined to represent a region where the corresponding admissible policy satisfies that the short-term reliability satisfaction event has an estimated probability not lower than a prescribed value at time \(t =t^{(c)}_{\mathrm{recover}}\).
Specifically, it is defined as a set of parameters $\big(\{\boldsymbol{f}_{i}^{(c, l)}, \boldsymbol{f}_{i,o}^{(c, l)}\}_{i\in\mathcal{V},c\in\mathcal{C},l\in \mathcal{L}},$ $\{\gamma_{\mathrm{long}}^{(c)}, \gamma_{\mathrm{short}}^{(c)},$ $T_{\mathrm{recover}}^{(c)}\}_{c\in\mathcal{C}}\big)$ such that 
$\hat{p}^{(c)}_{\mathrm{relia}}(t) \geq p^{(c)}_{\mathrm{resil}}, \forall c \in \mathcal{C}$ for the corresponding control policy in the admissible policy space given the commodities' characteristics, and the network configuration (including the outage situation)\footnote{The resilience region should be checked with fixed $n_{\mathrm{trial}}$, $T_{\mathrm{win}}$ and $p^{(c)}_{\mathrm{resil}}$ to ensure consistency of the region.}.
The estimation $\hat{p}^{(c)}_{\mathrm{relia}}(t)$ of the short-term reliability satisfaction probability $p^{(c)}_{\mathrm{relia}}(t)$ is calculated by the sample proportion---an unbiased estimator that estimates probability by the ratio of the number of event occurrences to the total number of trials.

\subsection{Problem Formulation}
The \emph{multi-commodity least-cost resilient and reliable network control} (MC-LC-ResRNC) problem aims at minimizing the total network cost while satisfying reliability and resilience constraints. 
Let the transmission and processing cost at time $t$ of commodity $c$ packets be defined as
\begin{align}
    &h^{(c)}(t) =\sum_{l\in\mathcal{L}^{(c)}}\sum_{(\imath,\jmath)\in\mathcal{E}^{(\phi_c)}}e_{\imath\jmath}\rho_{\imath\jmath}^{(\phi_c)} x_{\imath\jmath}^{(c, l)}(t),
\end{align}
where 
\begin{equation}
    e_{\imath\jmath} = \begin{cases}
    e_{ij} & \text{if }(\imath, \jmath) = (i_m, j_m) \\
    e_{i} & \text{if }(\imath, \jmath) = (i_m, i_{m+1}) \\
    \end{cases},
\end{equation} 
for some $m\in\mathcal{M}_{\phi_c}$.
% \begin{align}
%     &h^{(c)}(t) =\sum_{(i,j)\in\mathcal{E}}e_{ij}\underbrace{\Bigg(\sum_{l\in\mathcal{L}^{(c)}}\sum_{(\imath,\jmath)\in\mathcal{E}_{\mathrm{tr} , ij}^{(\phi_c)}}x_{\imath\jmath}^{(c, l)}(t) \Bigg)}_{\text{transmission flow on link } (i,j)}\nonumber\\
%     & \quad + \sum_{i\in\mathcal{V}}e_{i}\underbrace{\Bigg(\sum_{l\in\mathcal{L}^{(c)}}\sum_{(\imath,\jmath)\in\mathcal{E}_{\mathrm{pr} , i}^{(\phi_c)}}x_{\imath\jmath}^{(c, l)}(t)\Bigg)}_{\text{processing flow at node }i}.
% \end{align}
Then, the short-term cost over the time window $[t,t+T-1]$ is
\begin{align}
    &H^{(c)}(t,T)  =\frac{1}{T}\sum_{\tau=t}^{t+T-1}h^{(c)}(\tau).
\end{align}
We then formulate the following problem:
\vspace{-1mm}
\begin{subequations}
    \label{MC-LC-RNC}
    \begin{align}
    & \underset{\boldsymbol{x}(t)}{\mathrm{min}}
    & &  \lim_{T\rightarrow \infty}\sum_{c\in\mathcal{C}}\mathbb{E}\big\{H^{(c)}(0,T)\big\} \label{MC-LC-RNC_objective}\\
    & \;\mathrm{s.t.}
    & & 
    \eqref{queuing_dynamics}, 
    \eqref{pkt_expiration_and_consumption},
    \eqref{nonnegativity_constraint}, 
    \eqref{capacity_constraints},
    \eqref{availability_constraint}, \forall t \in \mathbb{N}\cup\{0\},\label{lifetime_based_queue_and_admissible_policy_space}\\
    &&& 
    \begin{aligned}
    &\lim_{T\rightarrow \infty }\mathbb{E}\{R^{(c)}(0, T)\}\\
    &\quad \geq  \gamma_{\mathrm{long}}^{(c)} \Xi_{\phi_{c}}^{(M_{\phi_c})}\lim_{T\rightarrow \infty }\mathbb{E}\{A^{(c)}(0, T)\},  \forall c\in \mathcal{C},
    \end{aligned} \label{MC-LC-RNC_reliability}
    \end{align}
\end{subequations}
\vspace{-5mm}
\begin{flalign}
    % \begin{aligned}
    %     &R^{(c)}(t - T_{\mathrm{win}}+1, T_{\mathrm{win}})\\
    %     & \geq \gamma_{\mathrm{short}}^{(c)}\Xi_{\phi_{c}}^{(M_{\phi_c})}A^{(c)}(t - T_{\mathrm{win}}+1, T_{\mathrm{win}}), \\
    %     &\qquad \forall t\in\mathcal{T}_{\mathrm{resil}}^{(c)}, c\in\mathcal{C},
    % \end{aligned}
    \begin{aligned}
      & p^{(c)}_{\mathrm{relia}}(t)\geq p^{(c)}_{\mathrm{resil}}, \\
      &\quad\forall t\in\mathcal{T}_{\mathrm{resil}}^{(c)}=\{t\in\mathbb{N}:t\geq t_{\mathrm{recover}}^{(c)}\},  c \in \mathcal{C}
    \end{aligned}\label{MC-LC-ResRNC_resilience}
\end{flalign}
where $\boldsymbol{x}(t)\in\mathbb{R}_+^{\sum_{c\in\mathcal{C}}L_{\mathrm{max}}^{(c)}|\mathcal{E}^{(\phi_c)}|}$ is the flow decision vector, which includes all the transmission and processing flow rate in the network, constraint \eqref{lifetime_based_queue_and_admissible_policy_space} includes the constraints for the rule of the lifetime-based queue and the constraints that form the admissible policy space, constraint \eqref{MC-LC-RNC_reliability} is the reliability constraint, and constraint \eqref{MC-LC-ResRNC_resilience} is the resilience constraint with $t^{(c)}_{\mathrm{recover}}=t_o+T^{(c)}_{\mathrm{recover}}$.

The \emph{Multi-Commodity Least-Cost Reliable Network Control} (MC-LC-RNC) problem is defined as problem~\eqref{MC-LC-RNC} without the resilience constraint~\eqref{MC-LC-ResRNC_resilience} and corresponds to the multi-commodity version of problem \(\mathscr{P}_0\) in~\cite{RCNC}. 
By incorporating resilience, we define the MC-LC-ResRNC problem as problem~\eqref{MC-LC-RNC}, subject to the resilience constraint~\eqref{MC-LC-ResRNC_resilience}.

\section{Resilient and Reliable Cloud Network Control (ResRCNC)}
When addressing the MC-LC-ResRNC problem, two issues hinder the exploitation of Lyapunov optimization, which has been widely used for network control \cite{tassiulas1990stability, Neely_book}.
One is the existence of the resilience constraint \eqref{MC-LC-ResRNC_resilience} in the MC-LC-ResRNC problem, where the short-term average is not allowed in Lyapunov optimization.
Another is the nonstandard lifetime-based queues and the dependence of flow variables on queue backlogs in \eqref{availability_constraint}, which are not amenable to standard Lyapunov techniques.
In this paper, we address the MC-LC-ResRNC problem by firstly avoiding the short-term average issue via removing the resilience constraint \eqref{MC-LC-ResRNC_resilience}, which reduces the formulation to that of the MC-LC-RNC problem, and secondly addressing the other issue by leveraging the structure of the Reliable Cloud Network Control (RCNC) algorithm we introduced in \cite{RCNC}. 
In this section, we first briefly introduce the multi-commodity reliable cloud network control (MC-RCNC) algorithm, which serves as the foundation. 
Then, the proposed multi-commodity resilient and reliable cloud network control (MC-ResRCNC) algorithm is described in detail, highlighting its improvements over the MC-RCNC algorithm.

\subsection{The Multi-commodity Reliable Cloud Network Control (MC-RCNC) Algorithm}

RCNC solves formulation $\mathscr{P}_{0}$ in \cite{RCNC}, which is the single-commodity version of the MC-LC-RNC problem without considering the processing of a service function.
The basic RCNC algorithm gives an admissible policy that satisfies the reliability constraint with a lifetime-based queue for the single-commodity scenario.
Furthermore, the extension of the RCNC algorithm with layered graphs to accommodate multiple commodities and computation, which we refer to as the Multi-Commodity RCNC (MC-RCNC) algorithm, is exactly a solution to the MC-LC-RNC problem.\footnote{The MC-RCNC algorithm is outlined in Appendix H in \cite{RCNC}.}
Given this state-of-the-art solution, we then solve the MC-LC-ResRNC problem based on the structure of the MC-RCNC algorithm with a targeted design to address the resilience requirements.
These adjustments significantly enhance the recovery speed of the timely throughput to a prescribed level after an outage while maintaining the timely throughput in the long run for reliable transmission for each commodity.
These results are demonstrated in Section \ref{sec:numerical_experiment} with comprehensive numerical experiments.

The RCNC algorithm operates as a two-way optimization method, using a virtual flow solution (the solution to a relaxed problem) to guide the actual flow solution toward the least-cost configuration in problem $\mathscr{P}_0$ in \cite{RCNC}.
The flow guidance, called \emph{flow matching}, is implemented through a linear program (LP) that minimizes the multi-slot Lyapunov drift of the virtual queue associated with the flow-matching constraint (referred to as the {\em request queue}) \cite{RCNC}. 
To enable effective flow matching, the capacities for the virtual flow, namely the virtual capacities, are adjusted at the end of every  {\em frame} (a fixed number of time slots) according to the request queue. 
The MC-RCNC algorithm, an extension of the RCNC algorithm, incorporates multiple commodities and their service functions by separately considering packets, queue backlogs, and their layered graphs for each commodity. 
With this structure, we introduce the proposed MC-ResRCNC algorithm in the following subsections.

\subsection{The Multi-Commodity Virtual Network Control Problem}
The \emph{multi-commodity virtual network control problem}, whose single-commodity version is the $\mathscr{P}_{2}$ in \cite{RCNC}, serves as the basis for the application of Lyapunov optimization techniques to the MC-LC-RNC problem. 
Analyzing $\mathscr{P}_{2}$ instead of $\mathscr{P}_{0}$ is motivated by the difficulty in addressing the unconventional queuing processes in the lifetime queues and the dependency of the flow decision on them.
It is formulated as
\begin{subequations}
    \label{virtual_problem}
    \begin{align}
    & \underset{\boldsymbol{\nu}(t)}{\mathrm{min}}
    & &  \lim_{T\rightarrow \infty}\sum_{c\in\mathcal{C}}H^{(c)}(0,T) \label{virtual_problem_objective}\\
    & \;\mathrm{s.t.}
    & & \eqref{nonnegativity_constraint}, \eqref{capacity_constraints}, \forall t \in \mathbb{N}\cup\{0\}, \text{ and } 
    \eqref{MC-LC-RNC_reliability}\\
    &&& 
    \begin{aligned}
    &\overline{\nu}_{\imath\to}^{(c,\geq l)}\leq \sum_{\jmath\in\delta_{\imath}^{(\phi_c, -)}} \zeta_{\jmath\imath}^{(\phi_c)}\overline{\nu}_{\jmath\imath}^{(c, \geq (l+1))}  \\
    & \quad + \overline{a}_{\imath}^{(c, \geq l)}, \;\forall \imath\in\mathcal{V}^{(\phi_c)} ,  l\in\mathcal{L}^{(c)}, c\in\mathcal{C},
    \end{aligned}
    \label{virtual_problem_causality}
    \end{align}    
\end{subequations}
where we denote the virtual flow rate  $\boldsymbol{\nu}(t)$ to distinguish it from the actual flow rate $\boldsymbol{x}(t)$ in the original problem \eqref{MC-LC-RNC}\footnote{The flow rates in the objective and all constraints mentioned in problem \eqref{virtual_problem}, which was denoted by the elements in $\boldsymbol{x}(t)$, are thus replaced by the corresponding ones in $\boldsymbol{\nu}(t)$.}, $\overline{z}=\lim_{T\to \infty}\frac{1}{T}\sum_{t=0}^{T-1}z(t)$ for any quantity $z$ in time slot $t$, and $z^{(\geq l)}(t)=\sum_{l^\prime\in\{\ell\in\mathcal{L}^{(\phi)}:\ell\geq l\}}z^{(l^\prime)}(t), z^{(\leq l)}(t)=\sum_{l^\prime\in\{\ell\in\mathcal{L}^{(\phi)}:\ell\leq l\}}z^{(l^\prime)}(t)$ for any lifetime-based quantity $z$. 
The constraint \eqref{virtual_problem_causality} is the causality constraint, which replaced the queue evolution constraints \eqref{queuing_dynamics},  \eqref{pkt_expiration_and_consumption} and the availability constraint \eqref{availability_constraint} in the MC-LC-RNC problem \eqref{MC-LC-RNC} by forcing the lifetime of all packets to decrease (by at least 1) on average as they traverse any node $\imath$ and ensuring the number of input packets of any node $\imath\in\mathcal{V}^{(\phi_c)}$ should exceed the one of the output packets on average.
This replacement eliminates the unconventional queuing process and the dependency of the flow decision on it, which are factors that fail Lyapunov optimization.
This relaxed problem \eqref{virtual_problem} can be solved through Lyapunov optimization via translating the average constraints \eqref{MC-LC-RNC_reliability} and \eqref{virtual_problem_causality} into the equivalent mean rate stability constraints of the corresponding virtual queues \cite{Neely_book}.
The virtual queues corresponding to the reliability constraint \eqref{MC-LC-RNC_reliability} and causality constraint \eqref{virtual_problem_causality} are
\begin{subequations}\label{virtual_queues}
\begin{flalign}
&\begin{aligned}
    &U_{d_{c}}^{(c)}(t+1) = \mathrm{max}\bigg\{0, \;  - \sum_{\imath\in\delta_{d_c}^{(\phi_c, -)}}\zeta_{\imath d_{c}}^{(\phi_c)}\nu_{\imath d_{c}}^{(c)}(t)\\
    & \quad +U_{d_{c}}^{(c)}(t) + \Xi_{\phi_c}^{(M_{\phi_c})}\gamma_{\mathrm{long}}^{(c)}\sum_{l\in\mathcal{L}_a^{(c)}}\sum_{i\in\mathcal{V}}a_{i}^{(c,l)}(t) 
    \bigg\},
\end{aligned}\label{virtual_queue_reliability}\\
&\begin{aligned}
    &U_{\imath}^{(c,l)}(t+1) = \mathrm{max}\bigg\{0, U_{\imath}^{(c, l)}(t) - a_{\imath}^{(\geq l)}(t)\\
    & \quad + \nu_{\imath \to}^{(c, \geq l)}(t)
    - \sum_{\jmath\in\delta_{\imath}^{(\phi_c, -)}}\zeta_{\jmath\imath }^{(\phi_c)}\nu_{\jmath\imath}^{(c, \geq (l+1))}(t)\bigg\},
\end{aligned}\label{virtual_queue_causality}
\end{flalign}
\end{subequations}
for all $\imath \in \mathcal{V}^{(\phi_c)}\backslash \{d_{c}\}$, respectively. 
The equivalent problem (the multi-commodity version of $\mathscr{P}_2^{e}$ in \cite{RCNC}) is thus 
\begin{subequations}
    \label{virtual_problem_e}
    \begin{align}
    & \underset{\boldsymbol{\nu}(t)}{\mathrm{min}}
    & &  \lim_{T\rightarrow \infty}\sum_{c\in\mathcal{C}}H^{(c)}(0,T) \label{virtual_problem_objective_e}\\
    & \mathrm{s.t.}
    & & \eqref{nonnegativity_constraint}, \eqref{capacity_constraints}, \forall t \in \mathbb{N}\cup\{0\}\\
    &&& \lim_{t\to\infty}\frac{\mathbb{E}\{U_{d_{c}}^{(c)}(t)\}}{t}=0, \forall c\in\mathcal{C}\\
    &&& \lim_{t\to\infty}\frac{\mathbb{E}\{U_{\imath}^{(c)}(t)\}}{t}=0, \forall \imath\in\mathcal{V}^{(\phi_c)}, c\in\mathcal{C},
    \end{align}
\end{subequations}
which can be solved by minimizing the \emph{Lyapunov drift-plus-penalty} with the \emph{max-weight algorithm}\cite{RCNC}, as described in the following. 
\subsection{Solution to the Multi-Commodity Virtual Network Control Problem}
\begin{figure*}[t]
    \begin{equation}\label{LDP}
        \begin{aligned}
            &\Delta (U(t)) 
            + V \sum_{c\in\mathcal{C}} h^{(c)}(t)
            % \leq B + V \sum_{c\in\mathcal{C}}\sum_{(\imath, \jmath)\in\mathcal{E}^{(\phi_c)}}\sum_{l\in\mathcal{L}^{(c)}}e_{\imath\jmath}\rho_{\imath\jmath}^{(\phi_c)}\nu_{\imath\jmath}^{(c)}(t) 
            % - \sum_{c\in\mathcal{C}}\beta_{d}^{(c)}U_{d_c}^{(c)}(t)\sum_{\imath\in\delta_{d_c}^{(\phi_c, -)}}\zeta_{\imath d_{c}}^{(\phi_c)}\sum_{l\in\mathcal{L}^{(c)}}\nu_{\imath d_{c}}^{(c, l)}(t) \\
            % &\quad - \sum_{c\in\mathcal{C}}\sum_{\imath\in\mathcal{V}^{(\phi_c)}}\sum_{l\in\mathcal{L}^{(c)}}\beta_{\imath}^{(c)} U_{\imath}^{(c, l)}(t)\left[\sum_{\jmath\in\delta_{\imath}^{(\phi_c, -)}}\zeta_{\jmath\imath}^{(\phi_c)}\nu_{\jmath\imath}^{(c, \geq (l+1))}(t)-\nu_{\imath \to}^{(c, \geq l)}\right] - \langle \tilde{\boldsymbol{a}}(t), \boldsymbol{\Sigma}\boldsymbol{U}(t)\rangle\\
            \leq B - \langle \tilde{\boldsymbol{a}}(t), \boldsymbol{\Sigma}\boldsymbol{U}(t)\rangle  \\
            & \quad - \sum_{c\in\mathcal{C}}\sum_{(\imath, \jmath)\in\mathcal{E}^{(\phi_c)}}\sum_{l\in\mathcal{L}^{(c)}}\underbrace{\left[-Ve_{\imath\jmath}-\frac{\beta_{\imath}^{(c)}U_{\imath}^{(c, \leq l)}(t)}{\rho_{\imath\jmath}^{(\phi_c)}} + \frac{\zeta_{\imath\jmath}^{(\phi_c)}\beta_{\jmath}^{(c)}}{\rho_{\imath\jmath}^{(\phi_c)}}\cdot \begin{cases}
                U_{d_c}^{(c)}(t), & \jmath = d_{c}\\
                U_{\jmath}^{(c,\leq (l-1))}(t), & \text{otherwise}
            \end{cases}\right]}_{w_{\imath\jmath}^{(c, l)}(t)}\underbrace{\left[\rho_{\imath\jmath}^{(\phi_c)}\nu_{\imath\jmath}^{(c,l)}(t)\right]}_{\tilde{\nu}_{\imath\jmath}^{(c, l)}(t)}
        \end{aligned}
    \end{equation}
\end{figure*}
To solve the problem \eqref{virtual_problem_e}, we minimize the Lyapunov drift-plus-penalty by minimizing its upper bound. 
The Lyapunov function is defined as $L(t)=\frac{1}{2}\|\Sigma\boldsymbol{U}(t)\|^2_2$, where $\boldsymbol{U}(t)$ is the vector that includes all virtual queues in \eqref{virtual_queues} for all $c\in\mathcal{C}$, and $\boldsymbol{\Sigma}$ is a diagonal matrix with diagonal elements being $\beta_{d}^{(c)}=1/\Xi_{\phi_c}^{(M_{\phi_c})}$ and $\beta_{\imath}^{(c,l)}=\beta_{\imath}^{(c)}=1/\Xi_{\phi_c}^{(m)}$ for $\imath=i_m$ corresponding to virtual queues in \eqref{virtual_queues}, respectively.
The Lyapunov drift is defined as $\Delta (U(t)) =L(t+1)-L(t)$.
The Lyapunov drift-plus-penalty along with its upper bound is thus given in \eqref{LDP}, where $B$ is the constant including the quadratic terms, $V$ is the fixed weight for the cost, and $\tilde{\boldsymbol{a}}(t)$ is a vector including $\gamma_{\mathrm{long}}^{(c)}\sum_{l\in\mathcal{L}^{(c)}}\sum_{i\in\mathcal{V}}a_{i}^{(c,l)}(t)$ and $a_{\imath}^{(\geq l)}(t)$, which also correspond to the virtual queues in \eqref{virtual_queues}, respectively.\footnote{The order of the diagonal elements in $\boldsymbol{\Sigma}$ and the elements in $\tilde{\boldsymbol{a}}(t)$ should align with those in $\boldsymbol{U}(t)$ according to the nodes $\imath\in\mathcal{V}^{(\phi_c)}$ the lifetimes $l\in\mathcal{L}^{(c)}$ and the commodities $c\in\mathcal{C}$.}

As a result, the minimization problem of the upper bound in \eqref{LDP} is equivalent to
\begin{subequations}
    \label{LDP_problem}
    \begin{align}
    & \underset{\tilde{\boldsymbol{\nu}}(t)}{\mathrm{max}}
    & &  \sum_{c\in\mathcal{C}}\sum_{(\imath, \jmath)\in\mathcal{E}^{(\phi_c)}}\sum_{l\in\mathcal{L}^{(c)}} w_{\imath\jmath}^{(c, l)}(t) \tilde{\nu}_{\imath\jmath}^{(c, l)}(t)\label{LDP_problem_objective}\\
    & \;\mathrm{s.t.}
    & & \eqref{nonnegativity_constraint}, \eqref{capacity_constraints}
    \end{align}    
\end{subequations}
The solution can be obtained by the max-weight algorithm as
\begin{equation}\label{max_weight_sol}
    \nu_{\imath\jmath}^{(c, l)}(t) = 
    \begin{cases}
        C_{ij}, & \text{if $(\imath, \jmath)=(i_m,j_m)$ for any $m$,}\\ 
        &\text{$(c, l, \imath, \jmath)=\underset{\substack{c\in\mathcal{C}, l\in\mathcal{L}^{(c)}, \\(\imath, \jmath)\in\mathcal{E}^{(\phi_c)}_{\mathrm{tr}, ij}}}{\mathrm{arg max}}w_{\imath\jmath}^{(c, l)}(t)$,} \\
        & \text{and $w_{\imath\jmath}^{(c, l)}(t)>0$}, \\
        C_{i}/\rho_{\imath\jmath}^{(\phi_c)}, & \text{if $(\imath, \jmath)=(i_m,i_{m+1})$ for any $m$,}\\ 
        &\text{$(c, l, \imath, \jmath)=\underset{\substack{c\in\mathcal{C}, l\in\mathcal{L}^{(c)}, \\(\imath, \jmath)\in\mathcal{E}^{(\phi_c)}_{\mathrm{pr}, i}}}{\mathrm{arg max}}w_{\imath\jmath}^{(c, l)}(t)$,}\\
        & \text{and $w_{\imath\jmath}^{(c, l)}(t)>0$}, \\
        0, & \text{otherwise}.
    \end{cases}
\end{equation}
\begin{remark}
    Choice of $V$: 
    The reasonable scale of $V$ can vary case by case due to the different configurations that result in different scales of the terms in weight $w_{\imath\jmath}^{(c, l)}(t)$. Specifically, it is determined by the scaling factors, workloads, and the virtual queue backlog sizes. We propose to approximate the scale by the following quantities:
    \begin{align}
        C_{\mathrm{avg}} &= \frac{1}{\sum_{c \in \mathcal{C}} |\mathcal{E}^{(\phi_c)}|}\sum_{c \in \mathcal{C}} \sum_{(\imath,\jmath) \in \mathcal{E}^{(\phi_c)}} \frac{C_{\imath\jmath}}{\rho_{\imath\jmath}^{(\phi_c)}}  \\
        e_{\mathrm{avg}} &= \frac{1}{\sum_{c \in \mathcal{C}} |\mathcal{E}^{(\phi_c)}|}\sum_{c \in \mathcal{C}} \sum_{(\imath,\jmath) \in \mathcal{E}^{(\phi_c)}} \frac{e_{\imath\jmath}\rho_{\imath\jmath}^{(\phi_c)}}{\beta_{\imath}^{(c)}}
    \end{align}
    The scale of $V$ can be approximated by $C_{\mathrm{avg}}/e_{\mathrm{avg}}$, where $C_{\mathrm{avg}}$ and $e_{\mathrm{avg}}$ are to approximate the virtual queue size and the coefficients in the weight, which is fair for a network with moderate traffic and balanced configuration. 
    Then, a reasonable $V$ can be chosen with $V=V'\times (C_{\mathrm{avg}}/e_{\mathrm{avg}})$. 
    The parameter $V'$ can be further chosen without changing its scale for different network configurations and traffic, and it is generally considered good for a value between 1 and 10.
\end{remark}

\subsection{Flow Matching}
With the solution to the multi-commodity virtual network control problem as the guidance, the actual solution to the MC-LC-RNC problem is then obtained through flow matching, which matches the actual flow assignment to the guiding virtual solution while satisfying all the constraints needed for an admissible control policy. 
Specifically, flow matching aims to satisfy the flow matching constraint 
\begin{equation}\label{flow_matching_constraint}
    \lim_{T\rightarrow \infty}\sum_{t=0}^{T-1} x_{\imath\jmath}^{(c, l)}(t) = \lim_{T\rightarrow \infty}\sum_{t=0}^{T-1} \nu_{\imath\jmath}^{(c, l)}(t)
\end{equation}
by stabilizing the \emph{request queue} through minimizing its multi-slot Lyapunov drift.

\subsubsection{Request Queue}
The request queues are defined as \cite{ResRCNC}
\begin{equation}\label{request_queue}
    R_{\imath\jmath}^{(c, l)}(t+1) = R_{\imath\jmath}^{(c, l)}(t) + \nu_{\imath\jmath}^{(c, l)}(t) -x_{\imath\jmath}^{(c, l)}(t)
\end{equation}
which is exactly the virtual queue \cite{Neely_book} of the flow matching constraint \eqref{flow_matching_constraint}.
Vector $\boldsymbol{R}(t)$ is defined to include all $R_{\imath\jmath}^{(c, l)}(t)$ in its elements. 

Note that the request queue proposed in \cite{RCNC} is in a different form, with the instantaneous virtual flow rate $\nu_{\imath\jmath}^{(c, l)}(t)$ replaced by its empirical average. 
Stabilizing the request queue in \cite{RCNC}, which is, in fact, the virtual queue of a variant of \eqref{flow_matching_constraint} that puts extra weight on the past values in virtual flow rates, can lead to an attachment to the historical flow decisions, and thus results in bad resilience due to its slow responsiveness to a sudden systematic change such as the network topology change in an outage event.

\subsubsection{Flow Matching Problem}
In the flow matching problem, we stabilize the request queue by minimizing its multi-slot Lyapunov drift, which is defined as 
\begin{equation}
    \Delta_n(\boldsymbol{R}(t))\triangleq \frac{\|\boldsymbol{R}(t+n-1)\|_2^2-\|\boldsymbol{R}(t)\|_2^2}{2}, 
\end{equation}
where $n$ is the number of look-ahead time slots. 
The upper bound of the multi-slot drift can be obtained through the telescope sum as
\begin{equation}\label{request_drift_upper_bound}
    \begin{aligned}
    \Delta_n&(\boldsymbol{R}(t))\leq B^\prime (t) \\
    & - \sum_{c\in\mathcal{C}}\sum_{(\imath,\jmath)\in\mathcal{E}^{(\phi_c)}} \sum_{l\in\mathcal{L}^{(c)}} \left\langle\boldsymbol{R}_{\imath\jmath}^{(c)}(t), \boldsymbol{M}_{\imath\jmath}^{(c)}(t)\mathbf{1}_{n}\right\rangle,
    \end{aligned}
\end{equation}
where $B^\prime (t)$ is a constant which includes all the quadratic terms, $\mathbf{1}_n\in\mathbb{B}^{n}$ is an all-one column vector,   $\boldsymbol{R}_{\imath\jmath}^{(c)}(t) = [R_{\imath\jmath}^{(c, 1)}(t) \dots R_{\imath\jmath}^{(c, L_{\mathrm{max}}^{(c)})}(t)]^T$, and 
\begin{equation*}
    \boldsymbol{M}_{\imath\jmath}^{(c)}(t) = 
    \begin{bsmallmatrix}
        x_{\imath\jmath}^{(c, 1)}(t) &x_{\imath\jmath}^{(c, 1)}(t+1) &\dots & x_{\imath\jmath}^{(c,1)}(t+n-1)\\
        x_{\imath\jmath}^{(c, 2)}(t) &x_{\imath\jmath}^{(c, 2)}(t+1) &\dots & x_{\imath\jmath}^{(c,2)}(t+n-1)\\
        \vdots & \vdots &\ddots & \vdots\\
        x_{\imath\jmath}^{(c,  L_{\mathrm{max}}^{(c)})}(t) &x_{\imath\jmath}^{(c,  L_{\mathrm{max}}^{(c)})}(t+1) &\dots & x_{\imath\jmath}^{(c,  L_{\mathrm{max}}^{(c)})}(t+n-1)
    \end{bsmallmatrix}.
\end{equation*}
Therefore, to minimize the multi-slot drift, the multi-commodity flow matching problem maximizes the last term in \eqref{request_drift_upper_bound} subject to the constraints required for an admissible policy. 
It is formulated as the following linear program (LP)
\begin{subequations}
    \label{flow_matching_problem}
    \begin{flalign}
    & \underset{\mathcal{M}(t)}{\mathrm{max}} 
    & &  \sum_{c\in\mathcal{C}}\sum_{(\imath,\jmath)\in\mathcal{E}^{(\phi_c)}} \sum_{l\in\mathcal{L}^{(c)}}\left\langle\boldsymbol{R}_{\imath\jmath}^{(c)}(t), \boldsymbol{M}_{\imath\jmath}^{(c)}(t)\mathbf{1}_{n}\right\rangle\label{flow_matching_objective}\\
    & \;\mathrm{s.t.}
    & & \sum_{c\in\mathcal{C}}\sum_{(\imath,\jmath) \in \mathcal{E}^{(\phi_c)}_{\mathrm{tr},ij}}\mathbf{1}_{L_{\mathrm{max}}^{(c)}}^T \boldsymbol{M}_{\imath\jmath}^{(c)}(t)\leq C_{ij}, \forall (i,j)\in\mathcal{E}\label{flow_matching_tr_capacity_constr}\\
    &&& \sum_{c\in\mathcal{C}}\sum_{(\imath,\jmath)\in \mathcal{E}^{(\phi_c)}_{\mathrm{pr},i}}\rho_{\imath\jmath}^{(\phi_c)}\mathbf{1}_{L_{\mathrm{max}}^{(c)}}^T \boldsymbol{M}_{\imath\jmath}^{(c)}(t)\leq C_{i}, \forall i\in\mathcal{V}\label{flow_matching_pr_capacity_constr}\\
    &&& 
    \begin{aligned}
        &\boldsymbol{D}_{L_{\mathrm{max}}^{(c)}}\sum_{\jmath\in\delta_{\imath}^{(\phi_c, -)}}g_{\tau}^{(c)}\big(\zeta_{\jmath\imath}^{(\phi_c)}\boldsymbol{M}_{\jmath\imath}^{(c)}(t)\big) \\
        &\geq \left[\sum_{\jmath\in\delta_{\imath}^{(\phi_c, +)}}g_{\tau+1}^{(c)}\big(\boldsymbol{M}_{\imath\jmath}^{(c)}(t)\big)\right]-g_{\tau+1}^{(c)}\big(\boldsymbol{A}_{\imath}^{(c)}(t)\big), \\
        & \quad\forall \imath\in\mathcal{V}^{(\phi_c)}, c\in\mathcal{C}, \tau\in\{0,1, \dots n-1\}
    \end{aligned}\label{flow_matching_availability_constr}\\
    &&& \boldsymbol{M}_{\imath\jmath}^{(c)}(t) \geq 0, \forall (\imath, \jmath)\in \mathcal{E}^{(\phi_c)}, c\in\mathcal{C},\label{flow_matching_nonnegativity_constr}
    \end{flalign}
\end{subequations}
where \eqref{flow_matching_tr_capacity_constr} is the transmission capacity constraint, \eqref{flow_matching_pr_capacity_constr} is the processing capacity constraint, \eqref{flow_matching_availability_constr} is the availability constraint, and \eqref{flow_matching_nonnegativity_constr} is the non-negativity constraint.
Set $\mathcal{M}(t)=\left\{\boldsymbol{M}_{\imath\jmath}^{(c)}(t): (\imath,\jmath)\in\mathcal{E}^{(\phi_c)}, c\in\mathcal{C}\right\}$ is the set of all $\boldsymbol{M}_{\imath\jmath}^{(c)}(t)$'s, which contain the actual flow rate of all lifetime in $n$ time slots ahead. 
The delay matrix $\boldsymbol{D}_{L_{\mathrm{max}}^{(c)}}\in\mathbb{B}^{L_{\mathrm{max}}^{(c)}\times L_{\mathrm{max}}^{(c)}}$ is an upper shift matrix to create a one-time-slot delay to its right-multiplied matrix.
The arrival matrix $\boldsymbol{A}_{\imath}^{(c)}(t)\in\mathbb{R}^{L_{\mathrm{max}}^{(c)}\times (n+1)}$ is defined with its $l$-th row being $\begin{bmatrix}Q_{\imath}^{(c, l)}(t) & \mathbf{1}_{n}^T\overline{a}_{\imath}^{(c, l)}(t)\end{bmatrix}$, where 
\vspace{-1mm}
\begin{flalign}
&\overline{a}_{\imath}^{(c, l)}(t)\hspace{-1mm}=\hspace{-1mm}
    \begin{cases}
      \frac{1}{T_{\mathrm{forget}}} \big[ (T_{\mathrm{forget}}-1) \,\overline{a}_{\imath}^{(c, l)}(t-1) + a_{\imath}^{(c, l)}(t) \big],\\
      \qquad\qquad\qquad\qquad\qquad\qquad\;\;\;\text{if } t\geq T_{\mathrm{forget}}\\
      \frac{1}{t+1} \big[ t \,\overline{a}_{\imath}^{(c, l)}(t-1) + a_{\imath}^{(c, l)}(t) \big], \,\text{if } t < T_{\mathrm{forget}}
    \end{cases}\label{empirical_avg_arrival}
\end{flalign}
is the empirical mean arrival rate with fading memory \cite{ResRCNC}.\footnote{We apply 1-based indexing when referring to the elements, columns, or rows in a matrix or a vector.}
Function $g_{\tau}^{(c)}:\mathbb{R}^{L_{\mathrm{max}}^{(c)}\times n}\to\mathbb{R}^{L_{\mathrm{max}}^{(c)}}$ is summing all columns of the input matrix with corresponding delays, which is defined as
\begin{equation}
    g_{\tau}^{(c)}(\boldsymbol{X})=\sum_{s=1}^{\tau} \boldsymbol{D}_{L_{\mathrm{max}}^{(c)}}^{\tau -s+1}\boldsymbol{X}[:, s]
\end{equation}
where $\boldsymbol{X}[:, s]$ is the $s$-th column of matrix $\boldsymbol{X}$.
The flow matching problem \eqref{flow_matching_problem}, as an LP, can be solved by standard solvers.
% \footnote{Applying a standard solver to this problem may require laborious reformulation involving highly sparse vectors and matrices. Therefore, a more robust solver that better handles corner cases is recommended.}

After obtaining the optimal $\boldsymbol{M}(t)$ by solving the flow matching problem \eqref{flow_matching_problem}, we can further get the desired actual flow rate decision for current time slot $t$ by collecting the first columns of $\boldsymbol{M}_{\imath\jmath}^{(c)}(t)$'s.
The rest of its columns are discarded, and the procedure repeats at the next time slot based on the updated information to make the corresponding decision. 
The MC-ResRCNC algorithm is thus presented as Algorithm \ref{alg:MC-ResRCNC}, where the virtual capacity update and the actions at outage time $t_o$, which are essential to the network resilience, are introduced in the following paragraphs and the next subsection.\footnote{Even though the algorithm involves the random variables in our description here, these steps can be applied with each realization of those random variables in each simulation, practical trial, or operation.}
\begin{algorithm}[t]
\caption{MC-ResRCNC algorithm}\label{alg:MC-ResRCNC}
\begin{algorithmic}[1]
    \For{frame $k=0,1,2,\dots$}
        \For{time slot $t=kK, kK+1,\dots, (k+1)K-1$}
        \If{$t = t_o$}
        \State Reset virtual capacity according to \eqref{reset_virtual_capacity};
        \State Redistribute frame-wise average virtual flow \Statex\hskip\algorithmicindent\hskip\algorithmicindent\hskip\algorithmicindent and actual flow according to \eqref{redistribute_avg};
        \EndIf
        \State Obtain the virtual flow from \eqref{max_weight_sol};
        \State Obtain the actual flow via solving \eqref{flow_matching_problem};
        \State Update the request queue according to \eqref{request_queue};
        \State Update the empirical average of mean arrival rates 
        \Statex\hskip\algorithmicindent\hskip\algorithmicindent with fading memory according to \eqref{empirical_avg_arrival};
        \EndFor
        \State Update the virtual capacities according to \eqref{virtual_capacity_update};
    \EndFor
\end{algorithmic}
\end{algorithm}

\subsubsection{Virtual Capacity Iteration} 
In the flow matching process, because the admissible policy space is a subset of the feasible policy space of the virtual problem, it is possible for the max-weight algorithm \eqref{max_weight_sol} to obtain an inadmissible policy, which violates the availability constraint \eqref{availability_constraint}, that is, transmission or processing with an unrealistically large flow that is beyond the number of available packets in the queue backlogs.
Such an inadmissible policy in the virtual network can cause a flow mismatch, leading to a degradation of system performance.
A straightforward solution to this issue, as mentioned in \cite{RCNC, ResRCNC}, is to reduce the \emph{virtual capacity} (i.e., the capacity in the virtual network) such that the actual flow rate is more likely to match the reduced virtual flow rate without violating the availability constraint \eqref{availability_constraint}.

Specifically, for all transmission and processing links, each virtual capacity is updated at the end of every \emph{frame}, a fixed number of time slots, according to 
\begin{subequations}\label{virtual_capacity_update}
    \begin{flalign}
        C_{ij}(k+1)&=C_{ij}(k)-\epsilon_{ij}(k)\mathbf{1}_{\{z\in\mathbb{R}:z> r_{\mathrm{min, tr}}\}}(\epsilon_{ij}(k)),
        \\
        C_{i}(k+1) &= C_{i}(k)-\epsilon_{i}(k)\mathbf{1}_{\{z\in\mathbb{R}:z>r_{\mathrm{min, pr}}\}}(\epsilon_{i}(k)),
    \end{flalign}
\end{subequations}
where capacities $C_{ij}(k)$ and $C_{i}(k)$ are the virtual capacities of transmission link $(i,j)$ and processing link at node $i$ at frame $k$, respectively. 
The indicator functions in \eqref{virtual_capacity_update} ensure that virtual capacities are updated only when the reduction values exceed their corresponding thresholds, thereby preventing unwanted updates due to the small flow difference caused by random arrivals.
The reduction values $\epsilon_{ij}(k)$ and $\epsilon_{i}(k)$ are calculated as
\begin{subequations}\label{capacity_reduction}
  \begin{flalign}
      &\epsilon_{ij}(k) = \sum_{c\in\mathcal{C}}\sum_{(\imath,\jmath)\in\mathcal{E}^{(\phi_c)}_{\mathrm{tr}, ij}}\rho_{\imath\jmath}^{(\phi_c)}\left[r_{\imath\jmath}^{(c)}(k)-r_{\to \imath}^{(c)}(k)\left(\bar{\nu}_{\imath\jmath}^{(c)}/\bar{\nu}_{\imath\to}^{(c)}\right)\right],\\
      & \epsilon_{i}(k) = \sum_{c\in\mathcal{C}}\sum_{(\imath,\jmath)\in\mathcal{E}^{(\phi_c)}_{\mathrm{pr}, i}}\rho_{\imath\jmath}^{(\phi_c)}\left[r_{\imath\jmath}^{(c)}(k)-r_{\to \imath}^{(c)}(k)\left(\bar{\nu}_{\imath\jmath}^{(c)}/\bar{\nu}_{\imath\to}^{(c)}\right)\right].
  \end{flalign}
\end{subequations}
The reduction values are determined based on $r_{\imath\jmath}^{(c)}(k)= \sum_{l\in\mathcal{L}^{(c)}} \max\Big\{0,\bar{\nu}_{\imath\jmath, k}^{(c, l)}-\bar{x}_{\imath\jmath, k}^{(c, l)}\Big\}$, which evaluate the flow difference with the frame-wise average virtual flow rate $\bar{\nu}_{\imath\jmath, k}^{(c, l)}= \bar{\nu}_{ij}^{(c,l)}((k+1)K-1)$ and the frame-wise average actual flow rate $\bar{x}_{\imath\jmath, k}^{(c, l)}= \bar{x}_{ij}^{(c,l)}((k+1)K-1)$.
They can be progressively calculated at each time slot in frame $k$ by
\begin{subequations}\label{frame_wise_avg}
    \begin{flalign}
    \bar{\nu}_{\imath\jmath}^{(c,l)}(t) &= \begin{cases}
        \nu_{\imath\jmath}^{(c,l)}(t), & \text{if } t = kK\\
        \frac{(t-kK)\,\bar{\nu}_{\imath\jmath}^{(c,l)}(t-1)+ \nu_{\imath\jmath}^{(c,l)}(t)}{t-kK+1}, & \text{otherwise}
    \end{cases}
    \\
    \bar{x}_{\imath\jmath}^{(c,l)}(t) &= \begin{cases}
        x_{\imath\jmath}^{(c,l)}(t), & \text{if } t = kK\\
        \frac{(t-kK)\,\bar{x}_{\imath\jmath}^{(c,l)}(t-1)+ x_{\imath\jmath}^{(c,l)}(t) }{t-kK+1}, & \text{otherwise}
    \end{cases}
    \end{flalign}
\end{subequations}
for all $(\imath,\jmath)\in\mathcal{E}^{(\phi_c)}$ and $c\in\mathcal{C}$.
The term $r_{\to \imath}^{(c)}(k)\left(\bar{\nu}_{\imath\jmath}^{(c)}/\bar{\nu}_{\imath\to}^{(c)}\right)$ in \eqref{capacity_reduction} is the insufficient input created by the capacity reduction on input links. It is subtracted here to avoid an overkill of capacity reduction \cite{RCNC}.
Note that the values of $r_{\mathrm{min, tr}}$ and $r_{\mathrm{min, pr}}$ can be manually chosen by checking numerical experiments' virtual capacity update step sizes when the flow matching is already achieved to prevent further reduction.

In comparison to the virtual capacity iteration in the MC-RCNC algorithm, the proposed one can achieve much more stable flow matching because the new design is based on the frame-wise flow difference with update thresholds. Moreover, finding the desired parameters (i.e., $r_{\mathrm{min, tr}}$, $r_{\mathrm{min, pr}}$) is easier in comparison to finding the $\kappa$ in \cite{RCNC}, which may take numerous simulations due to the lack of a clear guideline.

\subsection{Key Actions at Outage}
We assume that the occurrence of an outage event is communicated to the network controller (e.g., a software-defined network (SDN) controller) with negligible delay, which can be achieved through, e.g., short prioritized packets. 
Then, at outage time $t_{o}$, key resilience-specific procedures are included in the MC-ResRCNC algorithm: 1) The virtual capacities are reset to the actual capacities as
\begin{equation}\label{reset_virtual_capacity}
    C_{ij}(k) = C_{ij},\forall (i,j)\in\mathcal{E} \text{ and } C_{i}(k) = C_{i},\forall i\in\mathcal{V}
\end{equation}
to allow for potentially larger flows at non-outage links.
2) Before computing the frame-wise average flow at $t_o$ by \eqref{frame_wise_avg}, the one at $t_o-1$ is first redistributed as
\begin{subequations}\label{redistribute_avg}
    \begin{align}
        &\begin{aligned}
            &\hspace{-2.5mm}\bar{x}_{i_mj_m}^{(c, l)}(t_{o}-1) \gets \bar{x}_{i_mj_m}^{(c, l)}(t_{o}-1)\\
            &\hspace{-2.5mm}\;+ \underbrace{\frac{\bar{x}_{i_mj_m}^{(c, l)}(t_{o}-1)}{\sum_{(i^\prime,j^\prime)\in \mathcal{E}_{o} }\bar{x}_{i^\prime_m,j^\prime_m}^{(c, l)}(t_{o}-1)}}_{\substack{\text{percentage in the past flow on stage $m$}\\ \text{(exclude the flow on the outage links)}}} \underbrace{\sum_{(i^{''}_m,j^{''}_m)\in \mathcal{E}\backslash\mathcal{E}_{o}}\bar{x}_{i^{''}_mj^{''}_m}^{(c,l)}(t_{o}-1)}_{\text{all stage $m$ flows on the outage links}}\\
            & \forall m\in\mathcal{M}_{\phi_c}, \forall l\in\mathcal{L}^{(c)}, \forall c\in\mathcal{C}
        \end{aligned}\\
        &\begin{aligned}
            &\hspace{-2.5mm}\bar{x}_{i_mi_{m+1}}^{(c, l)}(t_{o}-1) \gets \bar{x}_{i_mi_{m+1}}^{(c, l)}(t_{o}-1)\\
            &\hspace{-2.5mm}\;+ \underbrace{\frac{\bar{x}_{i_mi_{m+1}}^{(c, l)}(t_{o}-1)}{\sum_{i^\prime\in \mathcal{V}_{o} }\bar{x}_{i^\prime_m,i^\prime_{m+1}}^{(c, l)}(t_{o}-1)}}_{\substack{\text{percentage in the past flow on stage $m$}\\ \text{(exclude the flow on the outage links)}}} \underbrace{\sum_{i^{''}_m\in \mathcal{V}\backslash\mathcal{V}_{o}}\bar{x}_{i^{''}_mi^{''}_{m+1}}^{(c,l)}(t_{o}-1)}_{\text{all stage $m$ flows on the outage links}}\\
            &\forall m\in\mathcal{M}_{\phi_c}\backslash\{M_{\phi_c}\}, \forall l\in\mathcal{L}^{(c)}, \forall c\in\mathcal{C}
        \end{aligned}
    \end{align}
\end{subequations}
to reduce the influence of the flow before outage on the frame-wise average flows $\bar{\nu}_{ij,k}^{(l)}$ and $\bar{x}_{ij,k}^{(l)}$.  
Finally, 3) the arrival rates after outage are adjusted to $\boldsymbol{\lambda}_{o}^{(c)}$ for each commodity $c$.

These key actions, which were not considered in the MC-RCNC algorithm, are crucial for resilience during an outage. 
It not only allows a fast reaction through capacity reset but also enables the virtual capacity iteration to gain a better understanding of the flow matching situation by redistributing the frame-wise average flow rates.
The arrival rate adjustment of the exogenous packet, which is part of the admission control rather than the routing/scheduling algorithm, is also vital for resilience, considering the potentially reduced reliability region after an outage, and will be applied in all numerical experiments with an outage event for fair comparisons in the next section.

% \textcolor{red}{=========== Chin-Wei's Progress Line ===========}

\section{Numerical Experiment}\label{sec:numerical_experiment}
% --- Abilene network topology ---
\begin{figure}[t]
    \centering \includegraphics[width=0.68\linewidth]{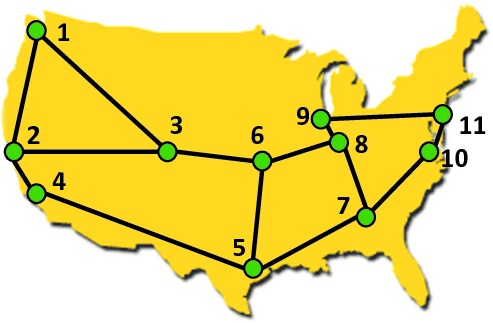}
    \vspace{-8pt}
    \caption{Abilene network topology (Reproduced from \cite{layered_graph} with permission)}
    \label{fig:Abilene_topology}
\end{figure}

In this section, we carry out the numerical experiments on the Abilene network topology shown in Fig. \ref{fig:Abilene_topology} to demonstrate the advantage of the MC-ResRCNC algorithm over the MC-RCNC algorithm and also show the behavior of MC-ResRCNC with different mean arrival rates after outage.
We follow the Abilene topology setting in \cite{DCNC}, where we use the dollar sign "\$" to indicate normalized cost units, “CPU” for processing resource units, and “bps” for transmission flow units.
Specifically, we assume that the network has bidirectional links with capacity 10 Gbps, cost 1 \$/Gbps for every transmission link, and capacity 20 CPU, cost 0.5 \$/CPU for every processing link.
We suppose that every transmission or processing over each link takes exactly one time slot, which is 14 ms, and every packet is 1 kbit long \cite{DCNC}.

We consider two typical services mentioned in \cite{DCNC}: a networking security service comprising a Deep Packet Inspection (DPI) function and an Internet Protocol Security (IPsec) function; and a video streaming service consisting of a Transrating function and a Transcoding function.
Their scaling factors and workloads (CPU/Mbps) are shown in Table \ref{table:service_parameters}.
\begin{table}[t]
\centering
\caption{Service parameters}
\label{table:service_parameters}
\renewcommand{\arraystretch}{1.1} % spacing between rows
\begin{tabular}{|c||cc|cc|}
\hline
\multirow{3}{*}{} & \multicolumn{2}{c|}{\textbf{Service 1 $(\phi=1)$}} & \multicolumn{2}{c|}{\textbf{Service 2 $(\phi=2)$}} \\
\cline{2-5}
 & \textbf{DPI} & \textbf{IPsec} & \textbf{Transrating} & \textbf{Transcoding} \\
 & $(m=1)$ & $(m=2)$ & $(m=1)$ & $(m=2)$ \\
\hline
$\xi_{\phi}^{(m)}$ & 1.0 & 2.3 & 1/3 & 1/2 \\
\hline
$r_{\phi}^{(m)}$ & 1/500 & 1/800 & 1/340 & 1/300 \\
\hline
\end{tabular}
\vspace{-15pt}
\end{table}
We consider six clients. Each of them forms one commodity with either one of the two services and a fixed source-destination pair. The details are shown in Table \ref{table:commodity_config}.
\begin{table}[t]
\centering
\caption{Commodity configuration}
\label{table:commodity_config}
\renewcommand{\arraystretch}{1.1}
\begin{tabular}{|c||c|c|}
\hline
\textbf{Commodity} $c$ & \textbf{Service} $\phi$ & (\textbf{Source}, \textbf{Destination}) \\
\hline
1 & 1 & (1, 7)\\
\hline
2 & 1 & (2, 10)\\
\hline
3 & 1 & (4, 11)\\
\hline
4 & 2 & (1, 7)\\
\hline
5 & 2 & (2, 10)\\
\hline
6 & 2 & (4, 11)\\
\hline
\end{tabular}
\end{table}
We suppose that the arrival of the exogenous packets follows a Poisson process, where, in each time slot $t$, the number of arrival packets $a_i^{(c, l)}(t)$ is Poisson distributed with parameter ($\lambda_i^{(c, l)}$).
Then, we impose the mean arrival rates before outage to be $\lambda_{s^{(c)}}^{(c, l)}=  1.64$\, Gbps for $c\in \{1, 2, 3\}$ and $\lambda_{s^{(c)}}^{(c, l)}= 2.46$\, Gbps for  $c\in \{4,5,6\}$ with $s^{(c)}$ being the corresponding source node for each commodity $c$ and $\mathcal{L}_a^{(c)} = \{6,7\}$ for all $c\in\mathcal{C}=\{1, 2, \dots, 6\}$. 
The mean arrival rates after the outage are assumed to be a fraction of the mean arrival rates before the outage.
All other $\lambda_i^{(c, l)}$ and $\lambda_{i,o}^{(c, l)}$ not mentioned above are zero. 
The required long-term reliability level is $\gamma_{\mathrm{long}}^{(c)}=90\%$ for all $c\in\mathcal{C}$.
We do the simulations with 100k time slot (1400 seconds) and the outage event happens at $t_{o}=45000$ (the end of the $630^{th}$ second), where node 6 and every associated link in Fig. \ref{fig:Abilene_topology} fail.
The number of Monte Carlo trials is $n_{\mathrm{trial}}=128$. 

Note that, for the ease of presentation, the ``long-term'' averages shown in the figures in this paper are all, in fact, the cumulative average from time slot $0$ to $t$, which will converge to the actual long-term averages as $t$ approaches infinity.
All lines in all plots are their sample mean $\hat{\mu}$ (ensemble mean) over the 128 trials, while the shadows shown in Fig. \ref{fig:compare_alg}, \ref{fig:diff_lambda_o_relia_level}, \ref{fig:diff_lambda_o_thrupt}, and \ref{fig:diff_lambda_o_cost} are their $\hat{\mu}\pm 1\hat{\sigma}$ region, where $\hat{\sigma}$ is the corresponding sample standard deviation.

\subsection{Comparison Between MC-ResRCNC and MC-RCNC}\label{sec:compare_ResRCNC_and_RCNC}
% --- Performance comparison between MC-RCNC and MC-ResRCNC ---
\begin{figure}[t]
\vspace{-7pt}
     \centering
     \subfloat[Average reliability level \label{fig:compare_alg_reliability}]{
         \centering
        \includegraphics[width=0.45\linewidth]{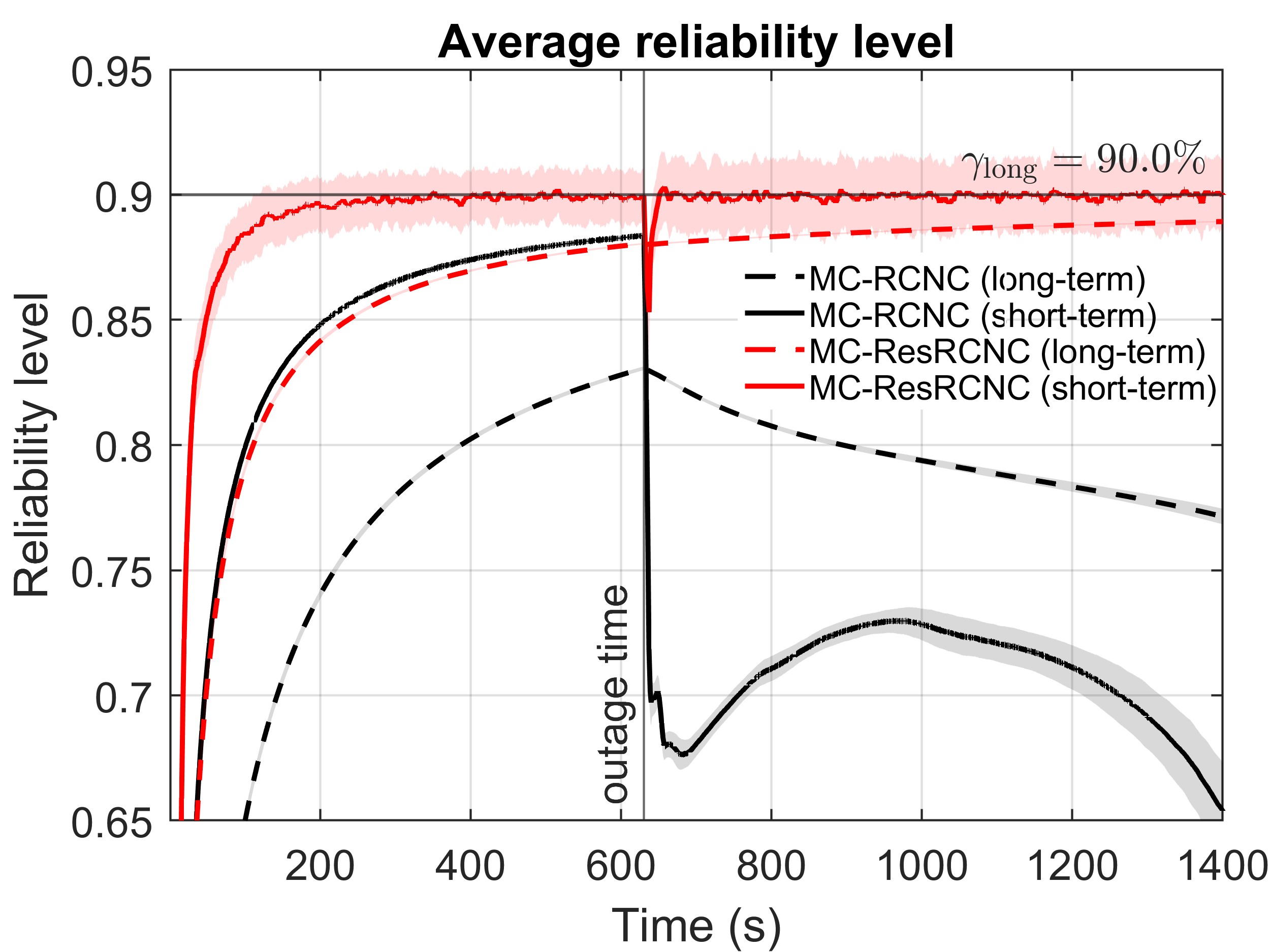}}
     \hfill
     \subfloat[Average cost \label{fig:compare_alg_cost}]{
         \centering
         \includegraphics[width=0.45\linewidth]{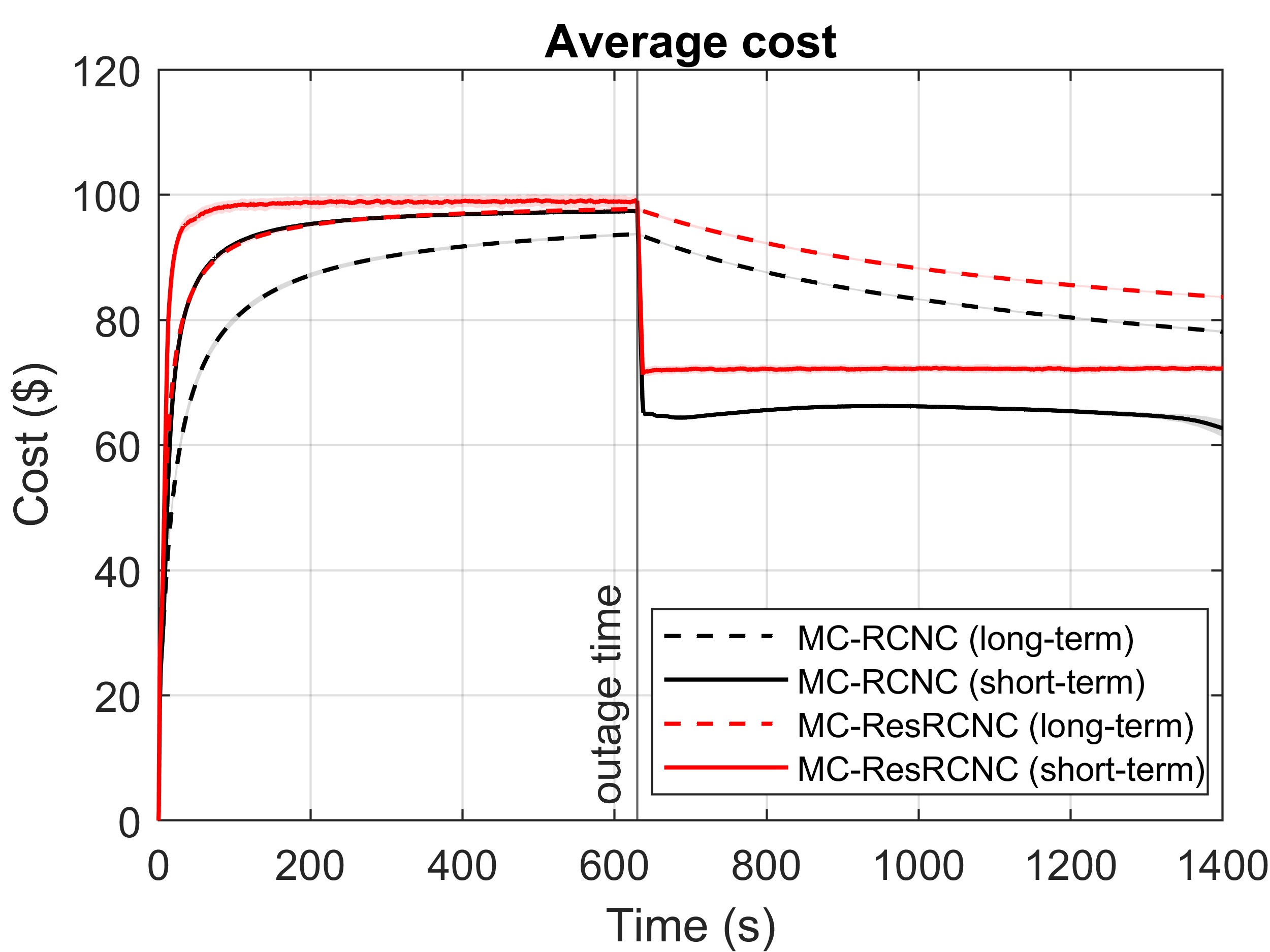}}
     \caption{Performance comparison between MC-RCNC and MC-ResRCNC}
\label{fig:compare_alg}
     \vspace{-15pt}
\end{figure}
% --- Flow matching plot (transmission) ---
\begin{figure}[t]
\vspace{-7pt}
     \centering
     \subfloat[MC-RCNC \label{fig:flow_match_RCNC_transm}]{
         \centering
         \includegraphics[width=0.45\linewidth]{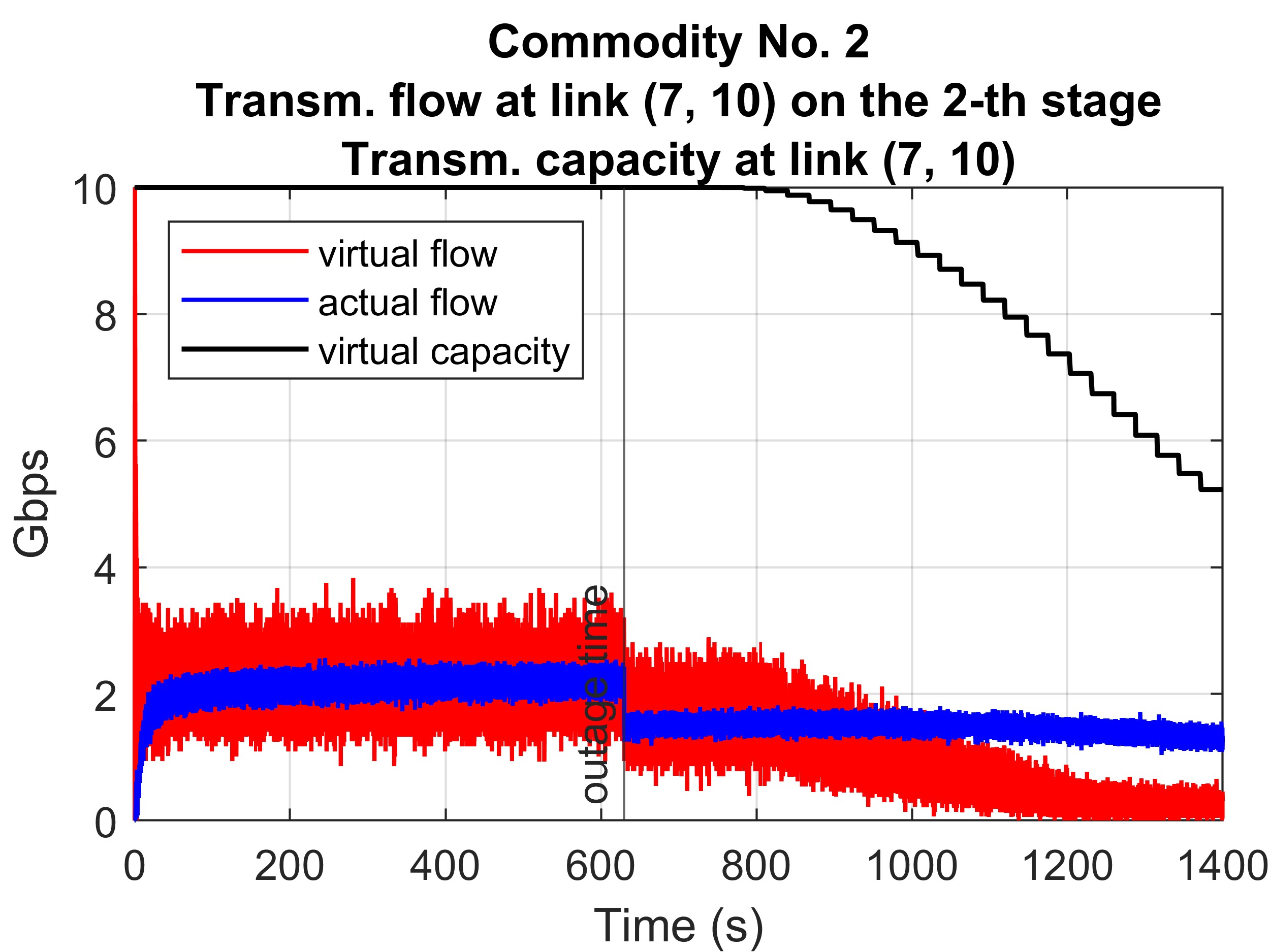}}
     \hfill
     \subfloat[MC-ResRCNC \label{fig:flow_match_ResRCNC_transm}]{
         \centering
        \includegraphics[width=0.45\linewidth]{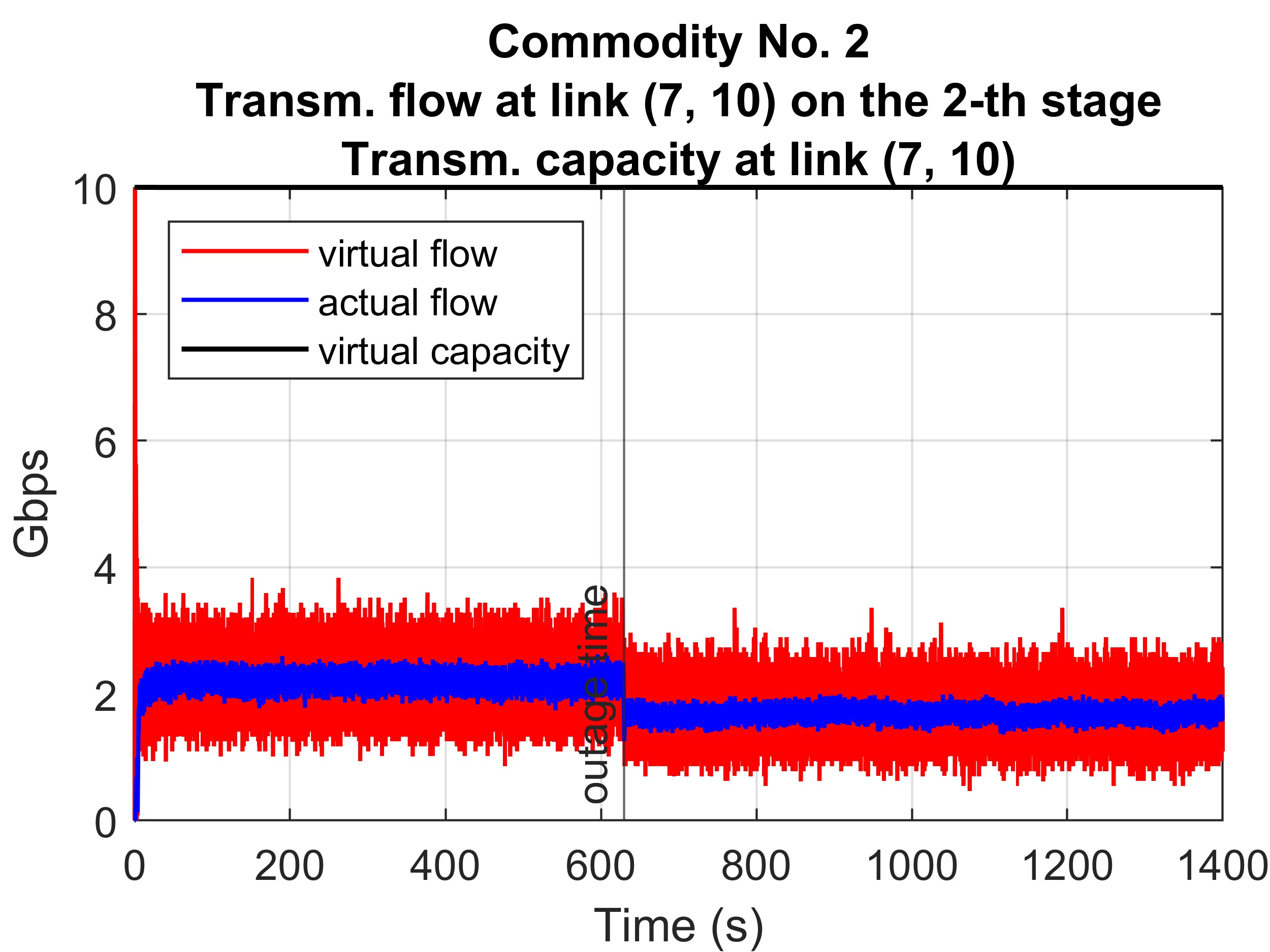}}
     \caption{Transmission flow matching comparison (Commodity 2, transmission link (7, 10) on stage 2)}
\label{fig:flow_match_transm}
     \vspace{-15pt}
\end{figure}
% --- Flow matching plot (processing) ---
\begin{figure}[t]
\vspace{-7pt}
     \centering
     \subfloat[MC-RCNC \label{fig:flow_match_RCNC_process}]{
         \centering
         \includegraphics[width=0.45\linewidth]{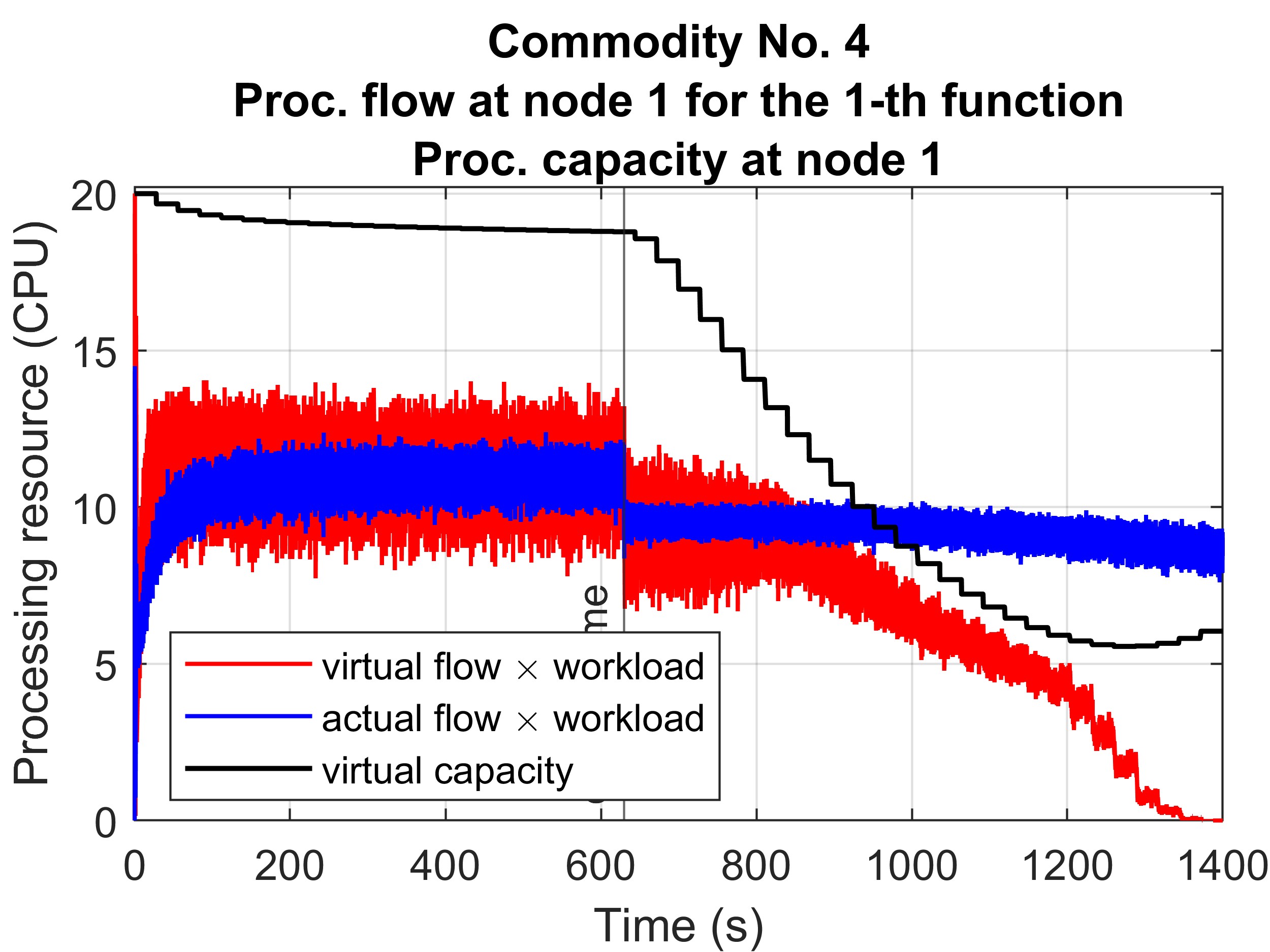}}
     \hfill
     \subfloat[MC-ResRCNC \label{fig:flow_match_ResRCNC_process}]{
         \centering
        \includegraphics[width=0.45\linewidth]{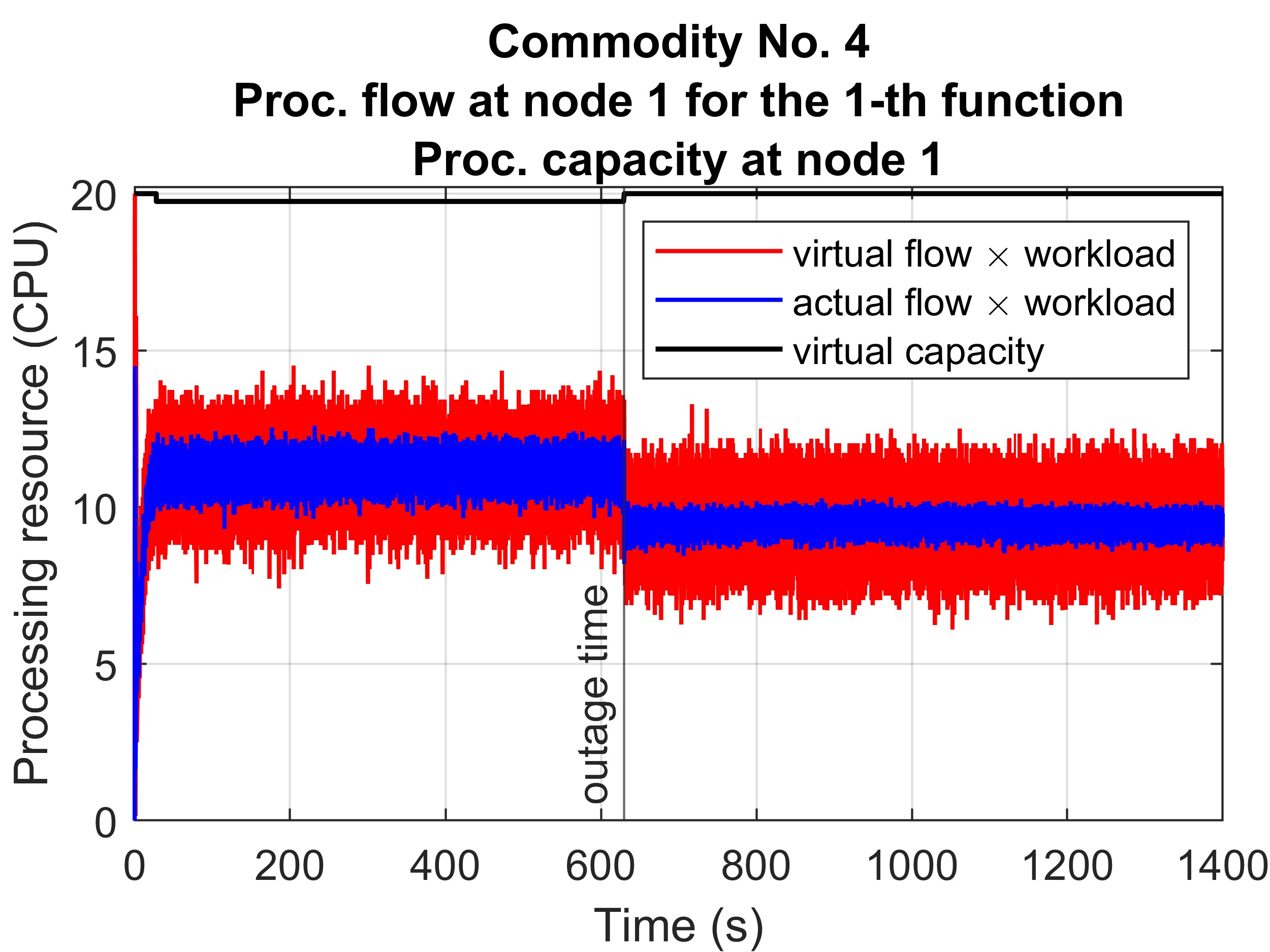}}
     \caption{Processing flow matching comparison (Commodity 4, function 1 (transrating) at node 1)}
\label{fig:flow_match_process}
     \vspace{-15pt}
\end{figure}
% --- Resilience region ---
\begin{figure}[t]
    \vspace{3pt}
    \centering \includegraphics[width=1\linewidth]{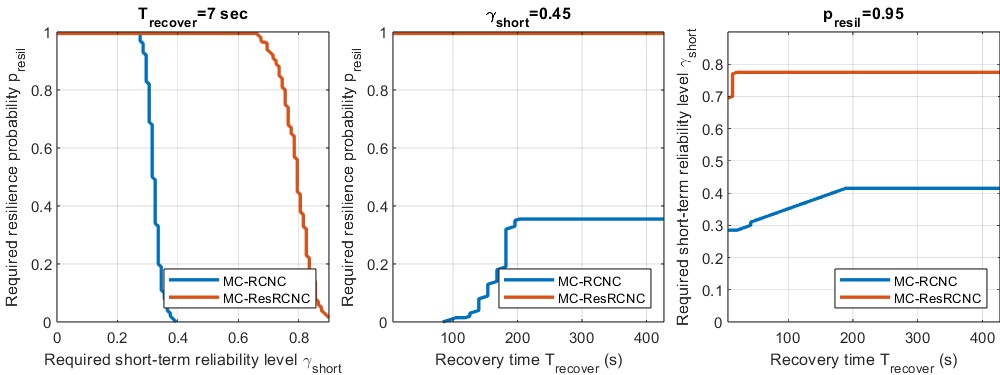}
    \vspace{-8pt}
    \caption{Resilience region}
    \label{fig:res_region}
    \vspace{-16pt}
\end{figure}
As we mentioned in the introduction, there does not yet exist a control algorithm that jointly considers reliability and resilience with a strict per-packet end-to-end latency constraint. 
Therefore, a good representative of the state-of-the-art algorithms in this scenario is our previously proposed MC-RCNC algorithm, which considers reliability with the strict per-packet end-to-end latency constraint.
We will thus compare the proposed MC-ResRCNC algorithm with the MC-RCNC algorithm throughout this subsection to show its advantages.

In the simulation, for both MC-ResRCNC and MC-RCNC, the frame size is $K=2000$ time slots, the weight for the cost in Lyapunov drift-plus-penalty is $V'=5$, the number of look-ahead time slots is $n=2$, the short-term average window size is $T_{\mathrm{win}}=500$, and the mean arrival rate after outage is the $0.75$ times the one before outage.
For MC-ResRCNC, the window size for fading memory is $T_{\mathrm{forget}}=500$, and the capacity update thresholds are $r_{\mathrm{min, tr}}=1000, r_{\mathrm{min, pr}}=0.1$. 
The parameter for capacity iteration in RCNC is $\kappa = 0.5$.

The performance comparison is demonstrated in Fig. \ref{fig:compare_alg}.
In Fig. \ref{fig:compare_alg_reliability}, both algorithms converge towards the 90\% reliability level before outage, but MC-ResRCNC converges much faster due to its faster adaptation to non-stationary events, including the start of operation. 
During an outage, the short-term reliability of MC-RCNC decreases significantly, and, instead of recovering monotonically, it is unstable over time, causing the long-term reliability to continue decreasing. 
On the other hand, the short-term reliability of MC-ResRCNC recovers quickly and remains at 90\% after the outage, resulting in a monotonically increasing long-term reliability towards 90\%.
As a result, we can see the significant advantage in both resilience and reliability of MC-ResRCNC over MC-RCNC.
In Fig. \ref{fig:compare_alg_cost}, the cost of MC-ResRCNC is generally higher than the cost of MC-RCNC. The reason is simply due to the generally lower throughput in MC-RCNC, as shown in Fig. \ref{fig:compare_alg_reliability}.

The stability comparison is demonstrated by the flow matching plots in Fig. \ref{fig:flow_match_transm} and \ref{fig:flow_match_process}.
Fig. \ref{fig:flow_match_transm} shows the instantaneous (transmission) virtual/actual flow rate with the corresponding virtual transmission capacity.
In Fig. \ref{fig:flow_match_RCNC_transm}, we can see the virtual flow and actual flow rate in MC-RCNC start to mismatch after the outage due to the overkill of virtual capacity update (the aggressively decreasing steps). 
In comparison, in Fig. \ref{fig:flow_match_ResRCNC_transm}, the flow rates in MC-ResRCNC stay matched before and after outage, which demonstrates the better operational stability in MC-ResRCNC compared to that in MC-RCNC.
A similar situation can also be observed in the processing flow rate times workload plots with the corresponding virtual processing capacity in Fig. \ref{fig:flow_match_process}.

Finally, the resilience region plots clearly show the better resilience of MC-ResRCNC compared to MC-RCNC based on the explicit definition in Section \ref{sec:res_region}.
In Fig. \ref{fig:res_region}, we fix each one of the 3 parameters $\gamma_{\mathrm{short}}, T_{\mathrm{recover}}$ and $p_{\mathrm{resil}}$\footnote{We set $\gamma_{\mathrm{short}}^{(c)}=\gamma_{\mathrm{short}}, T_{\mathrm{recover}}^{(c)}=T_{\mathrm{recover}}$, and $p_{\mathrm{resil}}^{(c)}=p_{\mathrm{resil}}$ for all $c\in\mathcal{C}$.} in each plot to draw the resilience region.
The curves show the boundaries of the corresponding resilience regions.
Obviously, all the plots in Fig. \ref{fig:res_region} show a larger resilience region in MC-ResRCNC, thereby demonstrating its advantage in resilience over MC-RCNC again.

\subsection{MC-ResRCNC with different arrival rates after outage}
% --- Reliability level of MC-ResRCNC with different lambda_o ---
\begin{figure}[t]
\vspace{-7pt}
     \centering
     \subfloat[Long-term reliability level (blue curve overlaps with red curve)\label{fig:diff_lambda_o_reliability_long}]{
         \centering
         \includegraphics[width=0.45\linewidth]{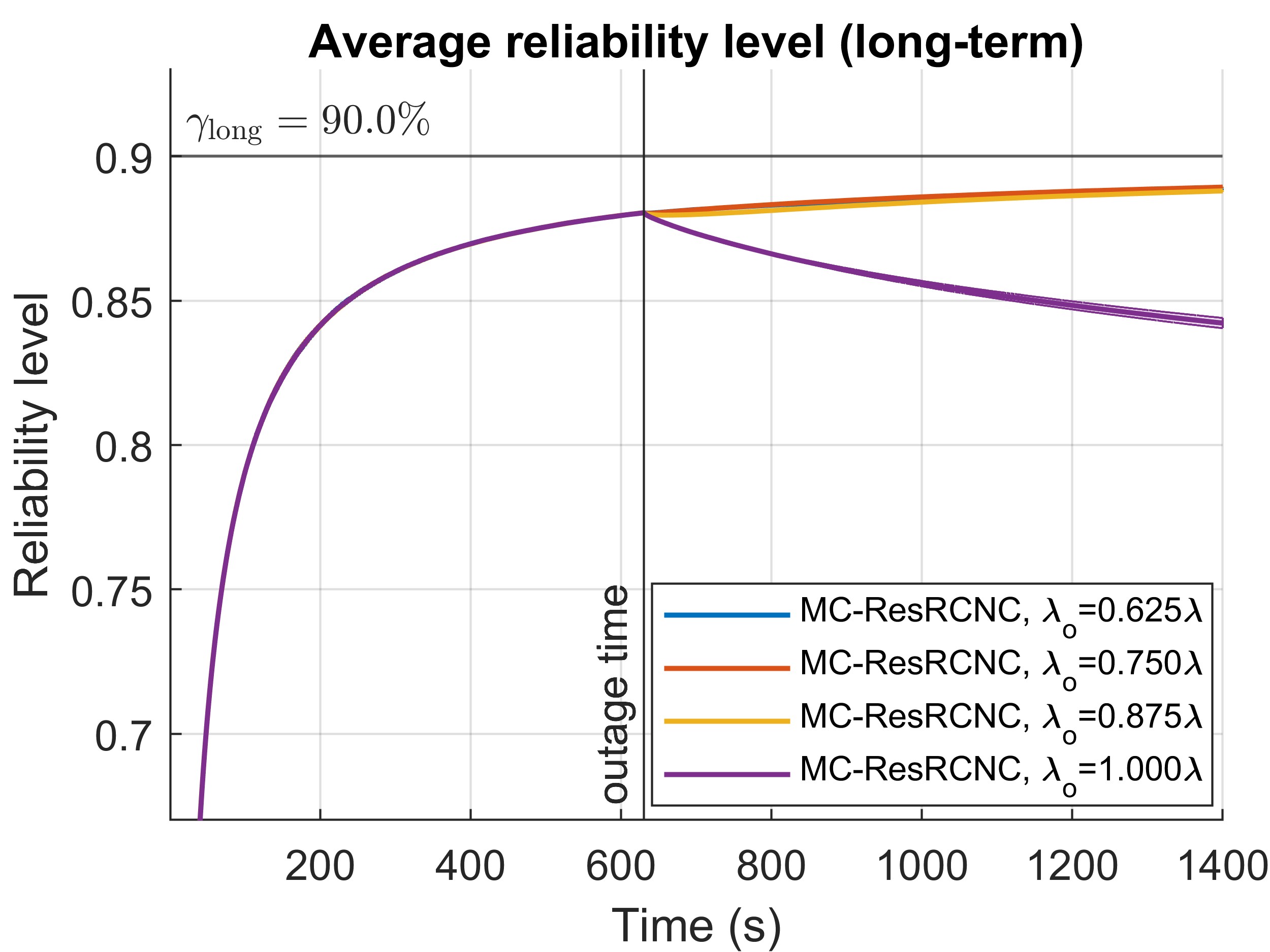}}
     \hfill
     \subfloat[Short-term reliability level \label{fig:diff_lambda_o_reliability_short}]{
         \centering
        \includegraphics[width=0.45\linewidth]{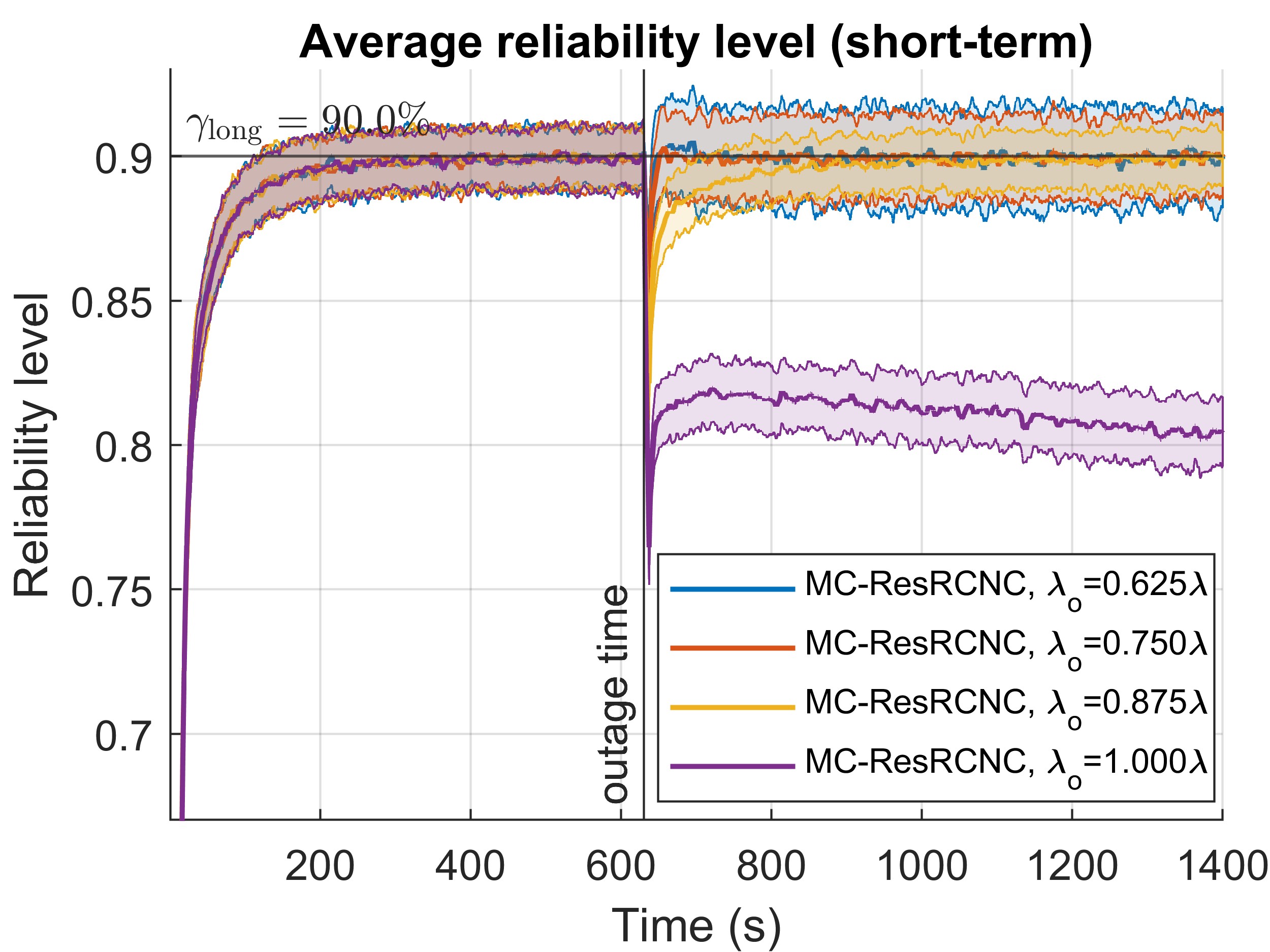}}
     \caption{Reliability level comparison with different mean arrival after outage}
\label{fig:diff_lambda_o_relia_level}
     \vspace{-15pt}
\end{figure}
% --- Timely throughput of MC-ResRCNC with different lambda_o ---
\begin{figure}[t]
\vspace{-7pt}
     \centering
     \subfloat[Long-term timely throughput \label{fig:diff_lambda_o_thrupt_long}]{
         \centering
         \includegraphics[width=0.45\linewidth]{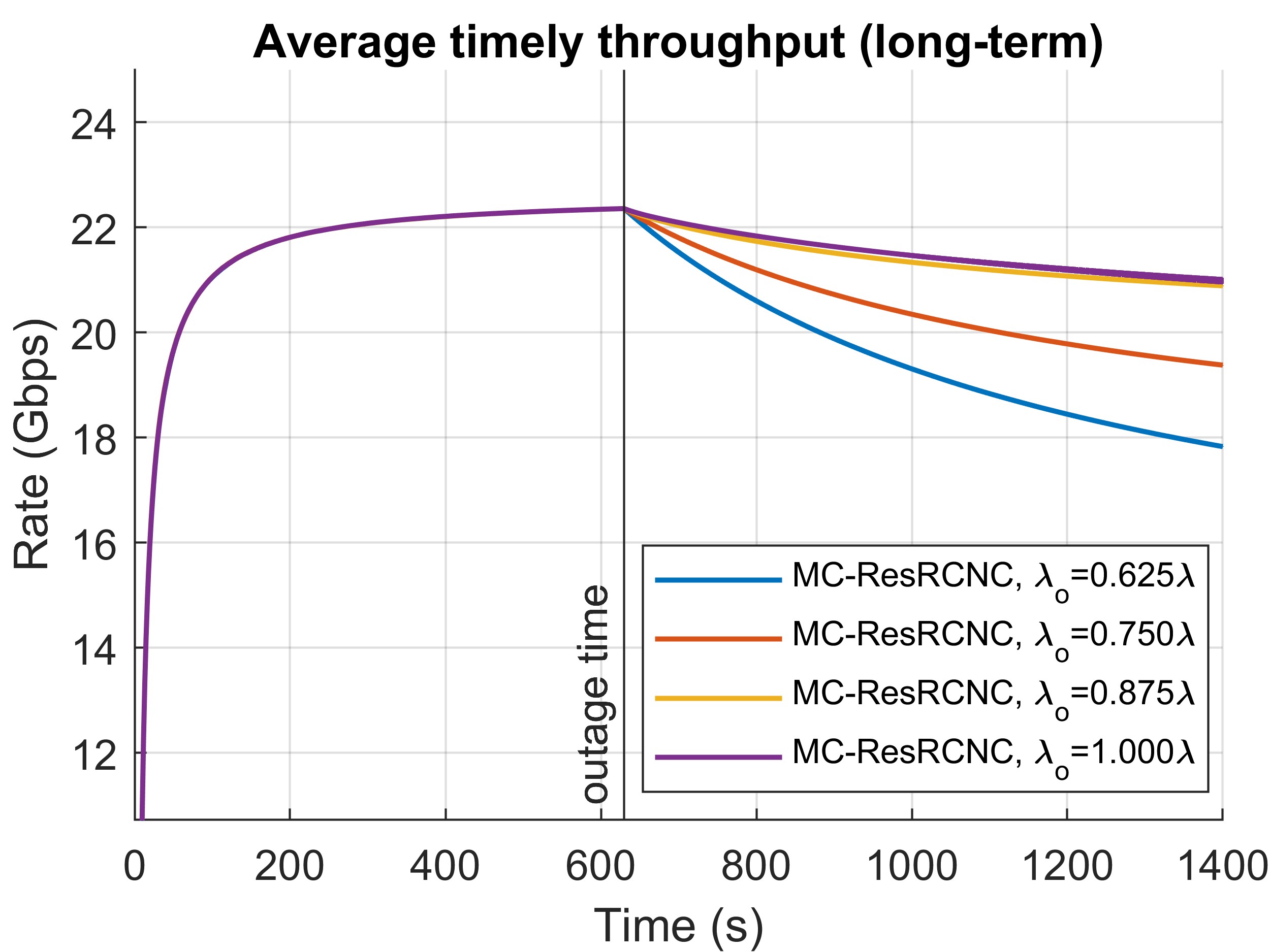}}
     \hfill
     \subfloat[Short-term timely throughput \label{fig:diff_lambda_o_thrupt_short}]{
         \centering
        \includegraphics[width=0.45\linewidth]{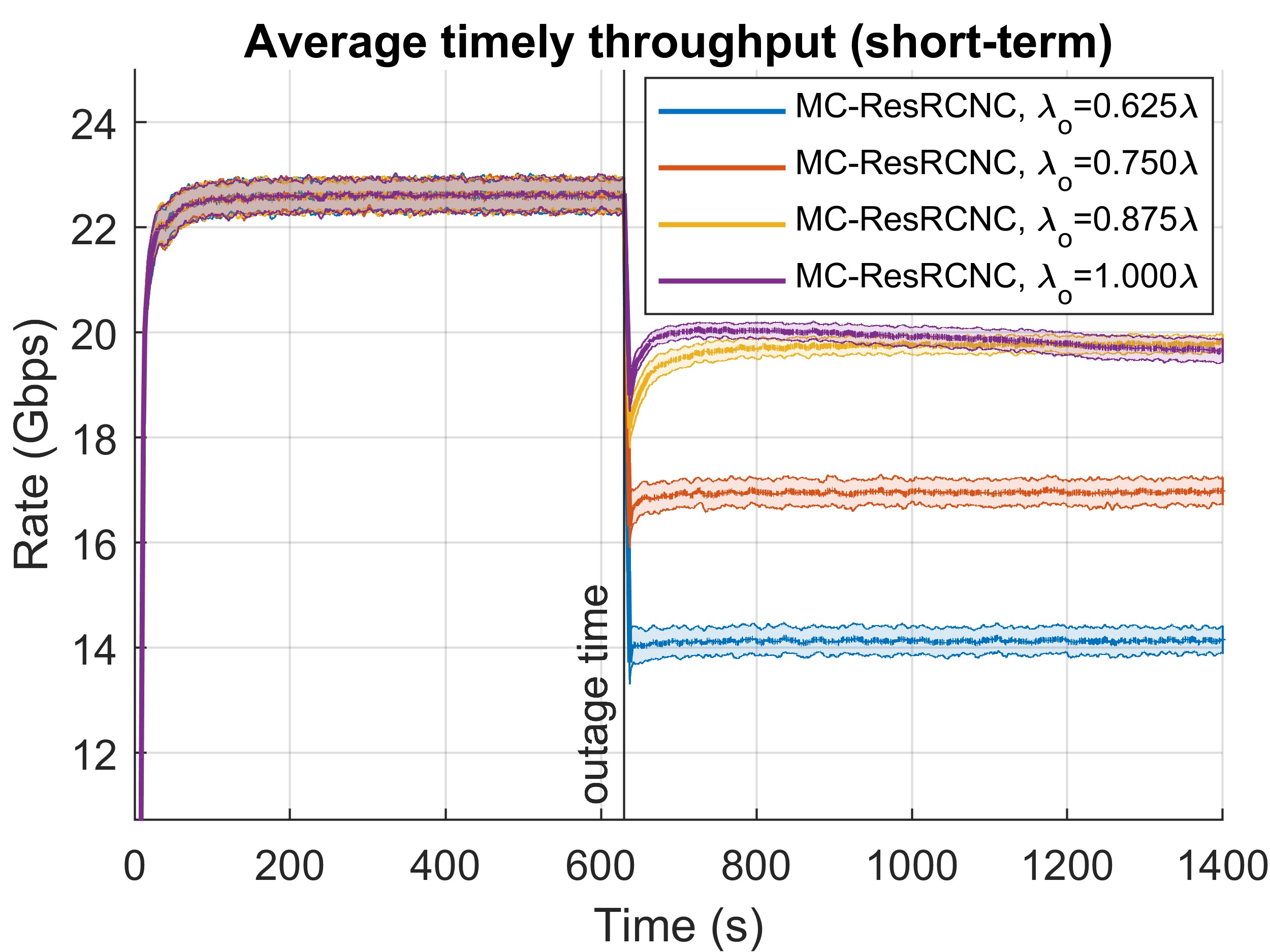}}
     \caption{Timely throughput comparison with different mean arrival after outage}
\label{fig:diff_lambda_o_thrupt}
     \vspace{-15pt}
\end{figure}
% --- Cost of MC-ResRCNC with different lambda_o ---
\begin{figure}[t]
\vspace{-7pt}
     \centering
     \subfloat[Long-term average cost \label{fig:diff_lambda_o_cost_long}]{
         \centering
         \includegraphics[width=0.45\linewidth]{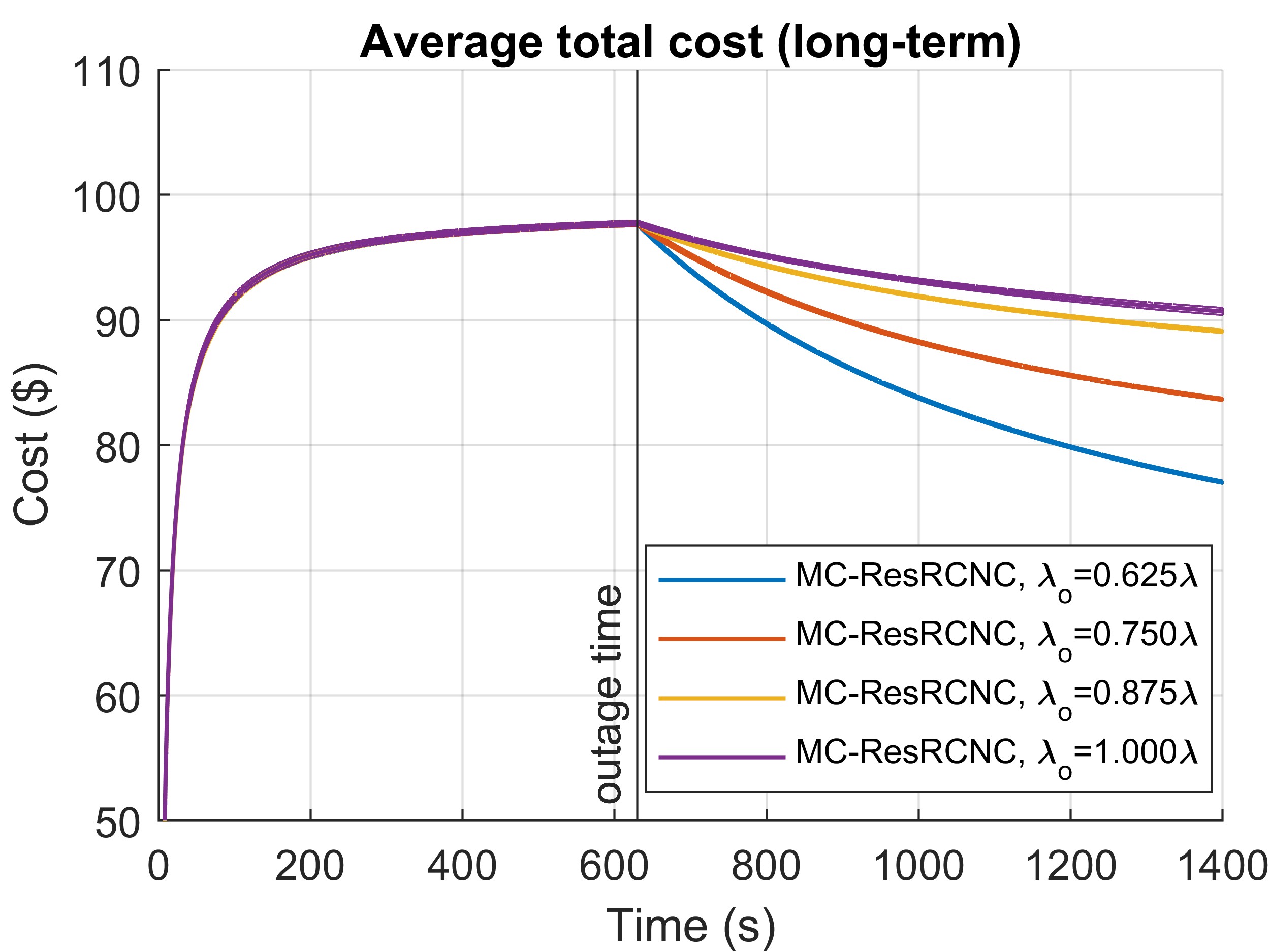}}
     \hfill
     \subfloat[Short-term average cost \label{fig:diff_lambda_o_cost_short}]{
         \centering
        \includegraphics[width=0.45\linewidth]{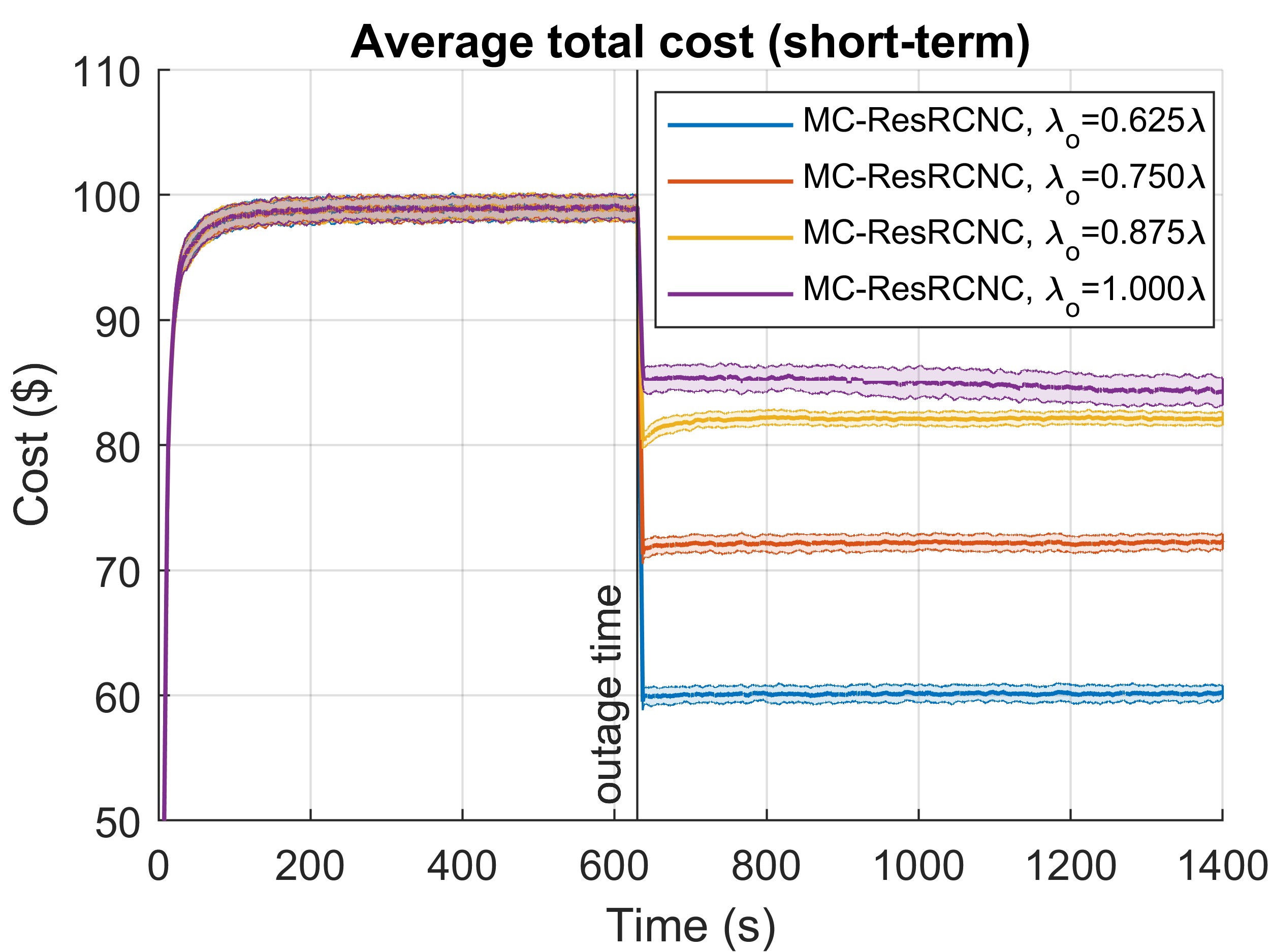}}
     \caption{Cost comparison with different mean arrival after outage}
\label{fig:diff_lambda_o_cost}
     \vspace{-15pt}
\end{figure}

In Section \ref{sec:outage_event}, we mentioned that the arrival rate after an outage should be reduced in response to the possible shrunken reliability region.  
A natural follow-up discussion is about how much it should be reduced. 
A too small $\boldsymbol{\lambda}_o^{(c)}$ can lead to a too conservative channel efficiency, while a too large $\boldsymbol{\lambda}_o^{(c)}$ can make the result outside the resilience region or even outside the reliability region, which implies the system never recovers. 
In this subsection, we demonstrate the behavior of MC-ResRCNC with different mean arrival rates after an outage by plotting the long-term and short-term averages of cost, timely throughput, and reliability level.
Specifically, we reuse all the simulation settings in Section \ref{sec:compare_ResRCNC_and_RCNC} except for the mean arrival rate after outage is $0.625, 0.750, 0.875,$ and $1.000$ times the mean arrival rate before outage.

In Fig. \ref{fig:diff_lambda_o_reliability_long}, we can see all cases converge to the 90\% requirement in reliability level except for the case with $\boldsymbol{\lambda}_o = 1.000 \boldsymbol{\lambda}$, which is outside the reliability region. 
Fig. \ref{fig:diff_lambda_o_reliability_short} shows that all the ensemble means of short-term reliability level can recover to the 90\% level except for the one in the case with $\boldsymbol{\lambda}_o = 1.000 \boldsymbol{\lambda}$, which indicates the maximum mean arrival rate of the reliability region after outage is between $0.875 \boldsymbol{\lambda}$ and $1.000 \boldsymbol{\lambda}$.
We denote the case with $\boldsymbol{\lambda}_o = 1.000 \boldsymbol{\lambda}$ as the ``unrecoverable case'' hereafter for the ease of presentation.
Even though all the other cases recover to 90\% in their mean, their variances differ.
Fig. \ref{fig:diff_lambda_o_reliability_short} demonstrates that the larger the mean arrival rate after outage is, the narrower the shadow is (except in the unrecoverable case).
The reason is probably that a heavier yet within-capacity traffic can lead to higher queue backlogs, which can alleviate the influence from random arrivals and thus lead to a more robust result.

Despite the unreliable and unresilient behavior of the unrecoverable case, in Fig. \ref{fig:diff_lambda_o_thrupt}, the magnitude of its timely throughput (both long-term and short-term) is surprisingly similar to the case with $\boldsymbol{\lambda}_o = 0.875 \boldsymbol{\lambda}$.
It might indicate that it is roughly the maximum available timely throughput after the outage. 
From the application side, the difference between the two cases is that if the exogenous rate is reduced by 12.5\%, then the {\em service} can decide which packets to drop to create an acceptable mean exogenous arrival rate that maintains the reliability level; in contrast, in the unrecoverable case, which packets are dropped is a consequence of random queue states. Thus, in most cases, the service quality in the latter case will be worse than in the former, even though the total number of packets arriving at the destination is similar. 

In Fig. \ref{fig:diff_lambda_o_cost}, both long-term and short-term plots show that a higher arrival rate leads to a higher cost, which is straightforward due to the corresponding higher throughput.
Furthermore, the case with $\boldsymbol{\lambda}_o = 0.875 \boldsymbol{\lambda}$ and the unrecoverable case have similar costs, which is also due to their similar timely throughput.

\section{Conclusion}
This paper introduces and formulates the multi-commodity least-cost resilient and reliable network control (MC-LC-ResRNC) problem with a specific definition of reliability and resilience constraints in terms of both long-term and short-term timely throughput metrics. 
A corresponding cloud network control algorithm---the multi-commodity resilient and reliable cloud network control (MC-ResRCNC) algorithm---is developed to solve the problem by leveraging Lyapunov optimization and flow matching techniques over the layered-graph structure with a targeted design for resilience.
Numerical experiments demonstrate the advantages of MC-ResRCNC in reliability, resilience, and operational stability. 
Analysis of the behavior of MC-ResRCNC after outage with different arrival rates reveals the necessity and possible strategy of reducing the arrival rate on the service side.
Future work will include designing distributed algorithms for large-scale distributed computing networks, as well as developing methods for lost packet recovery and rerouting based on network redundancy.

\section*{Acknowledgment}
This work is financially supported by NSF under the RINGS project 2148315. The authors would like to thank Dr. Hao Feng, Dr. Shilpa Talwar, and Dr. Rath Vannithamby (Intel) and Prof. Michael J. Neely (USC) for helpful discussions.

\bibliographystyle{IEEEtran}
\bibliography{IEEEabrv,bib_file}

\end{document}